\documentclass[aps,prc,twocolumn,showpacs,superscriptaddress,showpacs]{revtex4-1}
\usepackage{CJK}
\usepackage{graphicx}
\usepackage{dcolumn}
\usepackage{bm}
\usepackage{booktabs}
\usepackage{amsmath}
\usepackage{color}
\usepackage{array, longtable}
\usepackage{multirow}
\usepackage{hhline}
\usepackage{color,hyperref}
\hypersetup{colorlinks,breaklinks,linkcolor=blue,urlcolor=blue,anchorcolor=blue,citecolor=blue}
\begin{document}
\title{Chirality of $^{135}$Nd reexamined: Evidence for multiple chiral doublet bands}

\author{B.F. Lv}
 \affiliation{Centre de Sciences Nucl\'eaires et Sciences de la
  Mati\`ere, CNRS/IN2P3, Universit\'{e} Paris-Saclay, B\^at. 104-108, 91405  Orsay, France}
  \affiliation{Institute of Modern Physics, Chinese Academy of Sciences, Lanzhou 730000, China}
 \author{C. M. Petrache}
\affiliation{Centre de Sciences Nucl\'eaires et Sciences de la
  Mati\`ere, CNRS/IN2P3, Universit\'{e} Paris-Saclay, B\^at. 104-108, 91405  Orsay, France}
\author{Q. B. Chen}
\affiliation{Physik-Department, Technische Universit\"{a}t M\"{u}nchen, D-85747 Garching, Germany}
\author{J. Meng}
\affiliation{State Key Laboratory of Nuclear Physics and Technology, School of Physics, Peking University, Beijing 100871, China}
\affiliation{Yukawa Institute for Theoretical Physics, Kyoto University, Kyoto 606-8502, Japan} 
\author{A. Astier}
\affiliation{Centre de Sciences Nucl\'eaires et Sciences de la Mati\`ere, CNRS/IN2P3, Universit\'{e} Paris-Saclay, B\^at. 104-108, 91405  Orsay, France}
\author{E. Dupont}
 \affiliation{Centre de Sciences Nucl\'eaires et Sciences de la Mati\`ere, CNRS/IN2P3, Universit\'{e} Paris-Saclay, B\^at. 104-108, 91405  Orsay, France}
 \author{P. Greenlees}
 \affiliation{Department of Physics, University of Jyv\"askyl\"a, Jyv\"askyl\"a FIN-40014, Finland}
 \author{H. Badran}
\affiliation{Department of Physics, University of Jyv\"askyl\"a, Jyv\"askyl\"a FIN-40014, Finland}
\author{T. Calverley}  
 \affiliation{Department of Physics, University of Jyv\"askyl\"a, Jyv\"askyl\"a FIN-40014, Finland}
\affiliation{Department of Physics, University of Liverpool, The Oliver Lodge Laboratory, Liverpool L69 7ZE, United Kingdom}
\author{D. M. Cox}
\altaffiliation[Present address: ]{Department of Mathematical Physics, Lund Institute of Technology, S-22362 Lund, Sweden} 
 \affiliation{Department of Physics, University of Jyv\"askyl\"a, Jyv\"askyl\"a FIN-40014, Finland}
  \author{T. Grahn}
 \affiliation{Department of Physics, University of Jyv\"askyl\"a, Jyv\"askyl\"a FIN-40014, Finland}
 \author{J. Hilton}  
 \affiliation{Department of Physics, University of Jyv\"askyl\"a, Jyv\"askyl\"a FIN-40014, Finland}
 \affiliation{Department of Physics, University of Liverpool, The Oliver Lodge Laboratory, Liverpool L69 7ZE, United Kingdom}
\author{R. Julin}
 \affiliation{Department of Physics, University of Jyv\"askyl\"a, Jyv\"askyl\"a FIN-40014, Finland}
 \author{S. Juutinen}
  \affiliation{Department of Physics, University of Jyv\"askyl\"a, Jyv\"askyl\"a FIN-40014, Finland}
\author{J. Konki}
\altaffiliation[Present address: ]{CERN, CH-1211 Geneva 23, Switzerland} 
 \affiliation{Department of Physics, University of Jyv\"askyl\"a, Jyv\"askyl\"a FIN-40014, Finland}
\author{J. Pakarinen}
 \affiliation{Department of Physics, University of Jyv\"askyl\"a, Jyv\"askyl\"a FIN-40014, Finland}
\author{P. Papadakis}
\altaffiliation[Present address: ]{Oliver Lodge Laboratory, University of Liverpool, Liverpool L69 7ZE, United Kingdom} 
 \affiliation{Department of Physics, University of Jyv\"askyl\"a, Jyv\"askyl\"a FIN-40014, Finland}
\author{J. Partanen}
 \affiliation{Department of Physics, University of Jyv\"askyl\"a, Jyv\"askyl\"a FIN-40014, Finland}
\author{P. Rahkila}
 \affiliation{Department of Physics, University of Jyv\"askyl\"a, Jyv\"askyl\"a FIN-40014, Finland}
\author{P. Ruotsalainen}
 \affiliation{Department of Physics, University of Jyv\"askyl\"a, Jyv\"askyl\"a FIN-40014, Finland}
 \author{M. Sandzelius}
  \affiliation{Department of Physics, University of Jyv\"askyl\"a, Jyv\"askyl\"a FIN-40014, Finland}
\author{J. Saren}
 \affiliation{Department of Physics, University of Jyv\"askyl\"a, Jyv\"askyl\"a FIN-40014, Finland}
\author{C. Scholey}
 \affiliation{Department of Physics, University of Jyv\"askyl\"a, Jyv\"askyl\"a FIN-40014, Finland}
\author{J. Sorri}
 \affiliation{Department of Physics, University of Jyv\"askyl\"a, Jyv\"askyl\"a FIN-40014, Finland}
 \affiliation{Sodankyl\"a Geophysical Observatory, University of Oulu, FIN-99600 Sodankyl\"a, Finland} 
\author{S. Stolze}
\altaffiliation[Present address: ]{Physics Division, Argonne National Laboratory, Argonne, Illinois 60439, USA}
 \affiliation{Department of Physics, University of Jyv\"askyl\"a, Jyv\"askyl\"a FIN-40014, Finland}
 \author{J. Uusitalo} 
  \affiliation{Department of Physics, University of Jyv\"askyl\"a, Jyv\"askyl\"a FIN-40014, Finland}
 \author{B. Cederwall}
 \affiliation {KTH Department of Physics,S-10691 Stockholm, Sweden}
 \author{A. Ertoprak}
 \affiliation {KTH Department of Physics,S-10691 Stockholm, Sweden}
\author{H. Liu}  
 \affiliation {KTH Department of Physics,S-10691 Stockholm, Sweden}
\author{S. Guo}
\affiliation{Institute of Modern Physics, Chinese Academy of Sciences, Lanzhou 730000, China}
\author{M. L. Liu}
\affiliation{Institute of Modern Physics, Chinese Academy of Sciences, Lanzhou 730000, China}
\author{J. G. Wang}
\affiliation{Institute of Modern Physics, Chinese Academy of Sciences, Lanzhou 730000, China}
\author{X. H. Zhou}
\affiliation{Institute of Modern Physics, Chinese Academy of Sciences, Lanzhou 730000, China}
\author{I. Kuti}
\affiliation{Institute for Nuclear Research, Hungarian Academy of Sciences, Pf. 51, 4001 Debrecen, Hungary}
\author{J. Tim\'ar}
\affiliation{Institute for Nuclear Research, Hungarian Academy of Sciences, Pf. 51, 4001 Debrecen, Hungary}
\author{A. Tucholski}
\affiliation{University of Warsaw, Heavy Ion Laboratory, Pasteura 5a, 02-093 Warsaw, Poland}  
\author{J. Srebrny}
\address{University of Warsaw, Heavy Ion Laboratory, Pasteura 5a, 02-093 Warsaw, Poland}  
\author{C. Andreoiu}
\affiliation{Department of Chemistry, Simon Fraser University, Burnaby, BC V5A 1S6, Canada} 



\begin{abstract}

One new pair of positive-parity chiral doublet bands have been identified in the odd-$A$ nucleus $^{135}$Nd which together with the previously reported negative-parity chiral doublet bands constitute a third case of multiple chiral doublet (M$\chi$D) bands in the $A\approx130$ mass region. The properties of the M$\chi$D bands are  well reproduced by constrained covariant density functional theory and particle rotor model calculations. The newly observed M$\chi$D bands in $^{135}$Nd represents an important milestone in supporting the existence of M$\chi$D in nuclei. 
\end{abstract}



\maketitle


\section{Introduction}
The nuclei of the $A \approx 130$ mass region constitute the largest ensemble of chiral nuclei \cite{Stefan-Jie} in the table of elements \cite{adndt-chiral}. Recent reviews of both the experimental and theoretical investigation of the chiral bands can be found in Refs. \cite{frauendorf-2018,raduta-2016,meng-2016,meng-2014,Meng2010JPhysG37}. The first chiral bands were observed in odd-odd nuclei at low spins, where the shape can be $\gamma$-soft and the competition with other collective modes is important. At medium spins, the shape can change under the polarizing effect of unpaired nucleons resulting from broken pairs. In certain cases the triaxial shape becomes more rigid, being based on a deeper minimum of the potential energy surface. Three-quasiparticle chiral bands have been observed at medium spins in nuclei of the $A \approx 130$ mass region ($^{133}$La \cite{133la-petrache}, $^{133}$Ce \cite{Ayangeakaa2013Phys.Rev.Lett.172504}, $^{135}$Nd \cite{135nd-zhu}, $^{137}$Nd \cite{137nd-brant})  and in nuclei of the $A \approx 100$ mass region ($^{103}$Rh \cite{Kuti2014Phys.Rev.Lett.032501}, $^{105}$Rh \cite{Alcantara2004,105Rh}, $^{111,113}$Rh \cite{111-113Rh-luo}, $^{105}$Ag \cite{105Ag}, $^{107}$Ag \cite{107Ag}). Very recently, four- and six-quasiparticle chiral bands have also been observed in even-even nuclei of the $A \approx 130$ mass region ($^{136}$Nd \cite{136nd-PRCR} and $^{138}$Nd \cite{138nd-tilted}). In most nuclei only one chiral doublet was observed, but in a limited subset of chiral nuclei two or more chiral doublets were observed ($^{78}$Br \cite{Liu2016Phys.Rev.Lett.112501},  $^{103}$Rh  \cite{Kuti2014Phys.Rev.Lett.032501}, $^{105}$Rh  \cite{Alcantara2004}, $^{133}$Ce \cite{Ayangeakaa2013Phys.Rev.Lett.172504}, $^{136}$Nd \cite{136nd-PRCR}, $^{195}$Tl \cite{195Tl-roy}, and possibly $^{107}$Ag \cite{107ag-qi}),  exhibiting the exotic phenomenon of multiple chiral doublet bands (M$\chi$D)~\cite{Meng2006Phys.Rev.C037303,Peng2008Phys.Rev.C024309,Yao2009Phys.Rev.C067302,Li2011Phys.Rev.C037301,Droste2009EPJA, Chen2010PRC,Hamamoto2013PRC, Zhang2016CPC,136nd-qibo}. 
In the even-even nucleus $^{136}$Nd, the record of five nearly degenerate band pairs were identified, with properties in agreement with chiral interpretation within the tilted axis cranking constrained density functional theory (TAC-CDFT) framework \cite{136nd-PRCR}, which were also successfully described in detail  as chiral doublets in the framework of the multi-$j$ shell particle rotor model (PRM) \cite{136nd-qibo}.  
An alternative formalism predicting a new type of chiralitiy, different from that existing in odd-odd and odd-even nuclei, based on the generalized coherent state model, was also proposed and applied to the even-even nucleus  $^{138}$Nd \cite{raduta-jpg-2016,raduta-jpg-2017}.

\begin{figure*}[!htbp]
 \centering
\vskip -.cm
\rotatebox{-90}{\scalebox{0.35}{\includegraphics[width=\textwidth]{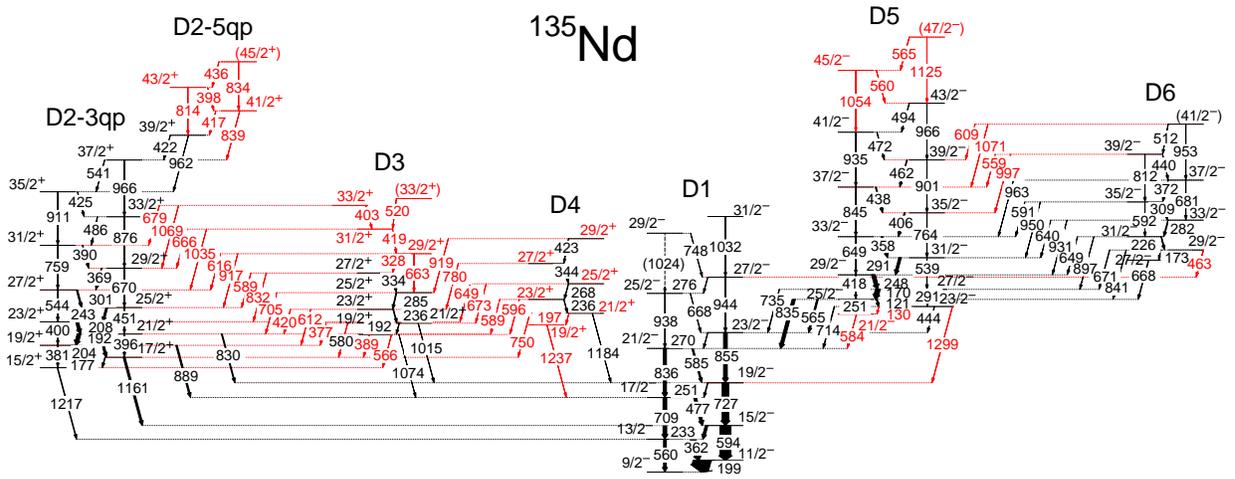}}}
\vskip .cm
\caption []{\label{fig1} (Color online) Partial level scheme of  $^{135}$Nd showing the newly identified doublet bands.}
\end{figure*}

The present work is devoted to the study of chirality in $^{135}$Nd, which is the isotone of $^{133}$Ce in which M$\chi$D bands have been recently reported \cite{Ayangeakaa2013Phys.Rev.Lett.172504}. We succeeded to identify one new chiral doublet band in $^{135}$Nd, composed of bands D3 and D4 , in addition to the previously know one composed of bands D5 and D6 (see Fig. \ref{fig1}), which raises to two the number of chiral doublet bands in this nucleus and gives a strong support to the existence of the M$\chi$D phenomenon in triaxial nuclei. The structure of the various bands is discussed within the constrained covariant density functional theory \cite{Meng2006Phys.Rev.C037303} and the PRM \cite{135nd-prm}.  
 Prior to this work, the chirality in $^{135}$Nd has been investigated both experimentally \cite{135nd-zhu,135nd-thesis,chiral-vibration} and theoretically \cite{135nd-prm,137nd-brant,135nd-pairing}.  

\section{Experimental results} 
\begin{figure*}[!htbp]
 \centering
\includegraphics[width=14.0cm]{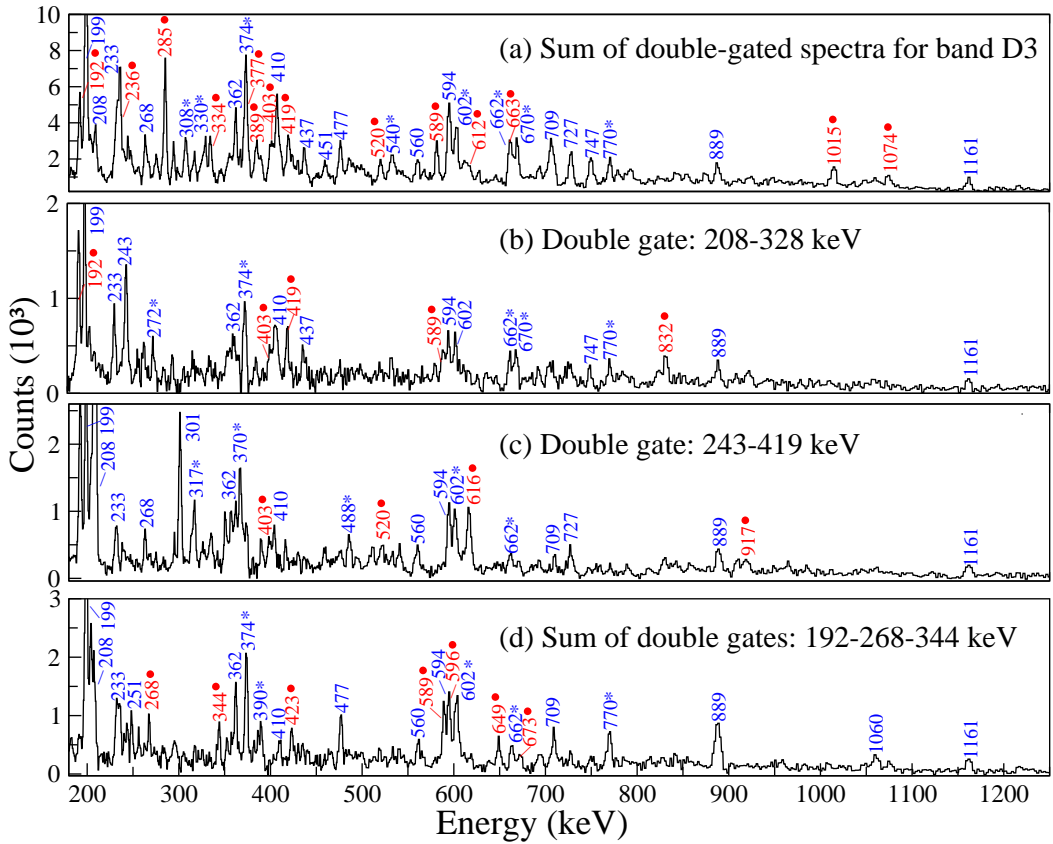}
\caption []{\label{fig2} (Color online) Spectra for the bands D3 and D4 of  $^{135}$Nd showing the newly identified transitions (in red color and marked with a dot).The peaks marked with an asterisk are contaminants from $^{136}$Nd.}
\end{figure*}
High-spin states in $^{135}$Nd were populated using the $^{100}$Mo($^{40}$Ar,5n) reaction at a beam energy of 152 MeV, provided by the K130 Cyclotron at the University of Jyv\"askyl\"a, Finland. We used as target a self-supporting enriched $^{100}$Mo foil of 0.5 mg/cm$^{2}$ thickness. The emitted $\gamma$-rays were detected by the JUROGAM II array, which comprised 39 Compton suppressed Ge detectors. Details of the experimental setup and data analysis can be found in Ref. \cite{lv}. The partial level scheme of $^{135}$Nd showing the previously known and the new chiral bands is given in Fig. \ref{fig1}. Spectra showing the newly identified transitions in the bands D3 and D4 are given in Fig. \ref{fig2}. Spin and parity assignments for newly observed levels were based on the measured directional correlation of oriented states ratios ($R_{\textrm{DCO}}$) and two-point angular correlation (anisotropy) ratios $R_{ac}$~\cite{lv,KramerFlecken1989333,Chiara.75.054305}. 
In this work, the definitions of $R_{\textrm{DCO}}$ and $R_{\textrm{ac}}$ are the same as those indicated in Ref.~\cite{lv}. 
In order to firmly establish the parity of bands D2-D4, we extracted the linear polarization of the $\gamma$ rays connecting the bands to the negative-parity band D1 following the procedure described in Ref. \cite{PhysRevC.92.044310}. The four transitions with energies of 830, 1015, 1161 and 1184 keV have IPDCO (integrated polarizational-directional correlation from oriented nuclei) \cite{STAROSTA199916} values of 0.043(8), 0.089(28), 0.008(2) and 0.018(5), respectively, which clearly show their electric character and therefore fix the positive-parity of bands D2, D3 and D4. Additionally, the detailed experimental information on the observed transitions is reported in Table.~\ref{135nd-table}.

Band D2-3qp, first reported in Ref.~\cite{135nd-piel}, is confirmed up to spin $39/2^+$. Three new levels are added on top of the band up to spin $(45/2^+)$, which together with the previously known $39/2^+$ level form a new band labelled D2-5qp. The spins of band D2 are established based on the dipole character of the several connecting transitions to band D1 (see Table.~\ref{135nd-table}). The positive parity of band D2 is experimentally established based on the electric character of the  830- and 1161-keV transitions. 

Band D3, previously known up to the level de-excited by the 334-keV transition \cite{135nd-thesis}, is extended by three more levels up to spin $(33/2^+)$. Another $33/2^+$ level decaying to both the $31/2^+$ level of band D3 and to band D2-3qp is also identified. Fourteen new transitions of 377, 389, 420, 566, 589, 612, 616, 666, 679, 705,  832, 917, 1035 and 1069 keV connecting band D3 to band D2-3qp are also identified. The parity of band D3 is changed to positive based on the $R_{\textrm{DCO}}$ and $R_{\textrm{ac}}$ values of the multitude of connecting transitions to band D1 and D2-3qp, in particular, those of the 566-, 705-, 832-, 1035-, and 1069-keV transitions which have $E2$ character (see Table.~\ref{135nd-table}), and the electric character of the 1015-keV transition.

Band D4, previously known up to the state de-excited by the 423-keV transition is confirmed \cite{135nd-thesis}, but based on $R_{\textrm{DCO}}$ values of the connecting transitions to bands D1 and D3, the spins are decreased by one unit and the parity is changed to positive. We add one new level with spin $19/2^+$ at the bottom of the band, one transition of 1237 keV towards band D1, four transitions of 596, 673, 780, and 919 keV towards band D3, and three transitions of 589, 649 and 750 keV towards band D2-3qp. The parity of band D4 is positive, similar to that of band D3, because the three connecting transitions of 589, 649, and 780 keV between bands D4 and D3 have firmly established E2 character, and the 1184-keV transition towards the negative-parity band D1 has E1 character (see Table.~\ref{135nd-table}).


We confirm all previously reported levels of bands D5 and D6 in Refs.~\cite{135nd-zhu, chiral-vibration}. Three new levels with spins 21/2$^-$, $45/2^-$ and $(47/2^-)$ are identified at the bottom and at the top of band D5, connected by the new transitions of 130, 560, 565, 1054, and 1125 keV. The two tentative transitions of 557 keV and 963 keV reported previously in Ref.~\cite{135nd-zhu} are confirmed, but our data show that the energy of the 557-keV transition is instead 559 keV. Three new transitions connecting band D6 to D5 with energies of 609, 997 and 1071 keV are newly identified. Three transitions of 463, 584, and 1299 keV from bands D5 and D6 to band D1 are also newly identified. 
 The spins and negative parity of band D5 are well established based on the E2 character of the 735-, 835-, and 1299-keV transitions towards band D1, deduced from the measured $R_{\textrm{DCO}}$ and $R_{\textrm{ac}}$ ratios (see Table.~\ref{135nd-table}).The spins and negative parity of band D6 are also well established based on the E2 character of the 897-, 931-, 950-, 963-, and 997-keV transitions towards band D5 (see Table.~\ref{135nd-table}).

\section{Discussion}

\begin{figure}[!htbp]
  \centering
  \vskip-0.5cm
 \includegraphics[width=8.5cm]{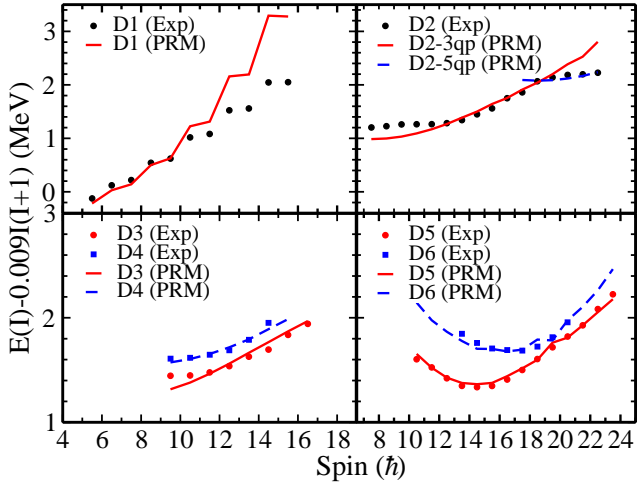}
  \caption{(Color online) Comparison between the experimental excitation energies relative to a reference rotor (symbols) and the particle rotor model calculations (lines) for the bands D1-D6. }
  \label{fig3}
\end{figure}

To understand the nature of the observed band structure in $^{135}$Nd, we can analyse the excitation energies of the bands, which reveal their detailed structure when drawn relative to a standard rotor reference, like in Fig. \ref{fig3}. One can observe a significant signature splitting between the two signatures of the yrast one-quasiparticle band built on the $\nu 9/2^-[514]$ Nilsson orbital, which indicates that $^{135}$Nd has a large triaxiality close to the ground state. One can also observe the parabolic behavior of the bands D2-D6, with a difference in excitation energy between the bands D2, D3 and D4 of around 200 keV,  and between the bands D5 and D6 of around 500 keV at low spins, which decreases steadily with increasing spin. Band D2 is observed over a much longer spin range than the bands D3, D4. It exhibits a markedly different slope at high spins, indicating a band crossing, and induced us to use the two labels D2-3qp and D2-5qp for the low- and high-spin parts, as they are interpreted as three (3qp) and five-quasiparticle (5qp) bands, respectively (see the discussion below). 

 \begin{figure}[]
 \vskip1.0cm
  \centering
    \includegraphics[width=8.5cm]{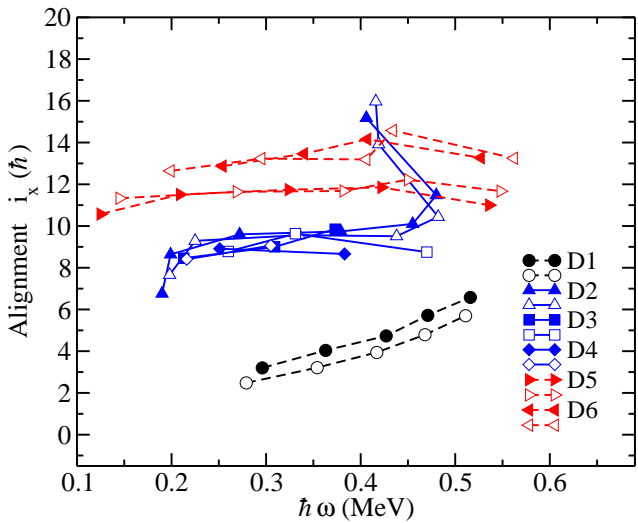}
  \caption{(Color online) The experimental quasi-particle alignments for the  chiral rotational bands of $^{135}$Nd.  The used Harris parameters are $\mathcal{J}_0 = 11$ $ \hbar^2$MeV$^{-1}$ and $\mathcal{J}_1=20$ $\hbar^4$MeV$^{-3}$.}
  \label{fig4}
\end{figure}

Another quantity which reveals the properties of the bands is the single-particle alignment $i_x$, which is given in Fig. \ref{fig4}. The used Harris parameters $\mathcal{J}_0 = 11$ $ \hbar^2$MeV$^{-1}$ and $\mathcal{J}_1=20$ $\hbar^4$MeV$^{-3}$ induce a flat behavior of band D2-D4 in the medium spin range, difference from that of band D1 which is up-sloping with increasing rotational frequency. This clearly shows the polarizing effect on the core of the two additional nucleons contributing to the 3qp configurations of bands D2-D4, which induces larger deformations. Particular features are exhibited by band D2 at both low and high spins. At the lowests spins one observes a decrease of $i_x$, most probably induced by the interaction with other positive-parity levels not shown in the partial level scheme given in Fig. \ref{fig1}, which will be published in a separate paper \cite{135nd-wobb}. At high spins one observes a backbending with a spin gain of at least  7$\hbar$, indicating the alignment of two more nucleons, most probably protons occupying the $h_{11/2}$ orbital. Furthermore, the very similar values of around 9$\hbar$ exhibited by the bands D2-D4, suggest similar 3qp configurations. The difference of $\approx 7 \hbar$ at low frequency between the bands D2-D4 and D1, strongly suggests the involvement of two more nucleons placed on opposite parity orbitals in the bands D2-D4, most probably protons occupying the $h_{11/2}$ and the strongly mixed $(d_{5/2},g_{7/2})$ orbitals.   
The single-particle alignment $i_x$ of the bands D5 is larger by 2-3$\hbar$ than that of the bands D2-D4, while that of band D6 is larger than that of band D5 by $\approx$ 2$\hbar$. The higher spin alignment of the bands D5, D6 relative to that of bands D2-D4 clearly shows the involvement of a pair of protons in the $h_{11/2}$ orbital.

\begin{figure}[htb]
  \centering
 \includegraphics[width=8.5cm]{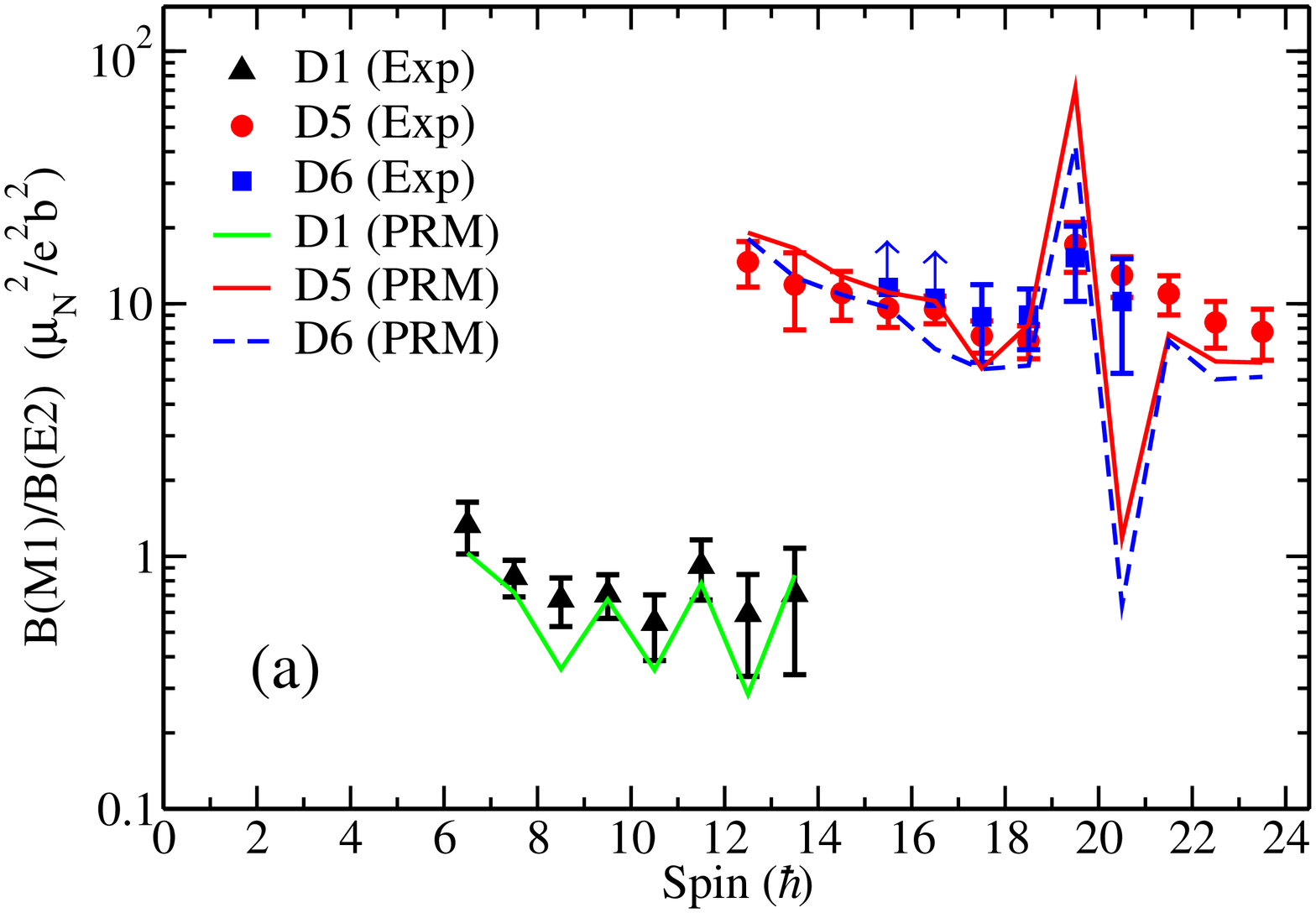}
 \vskip 1.0 cm
 \includegraphics[width=8.5cm]{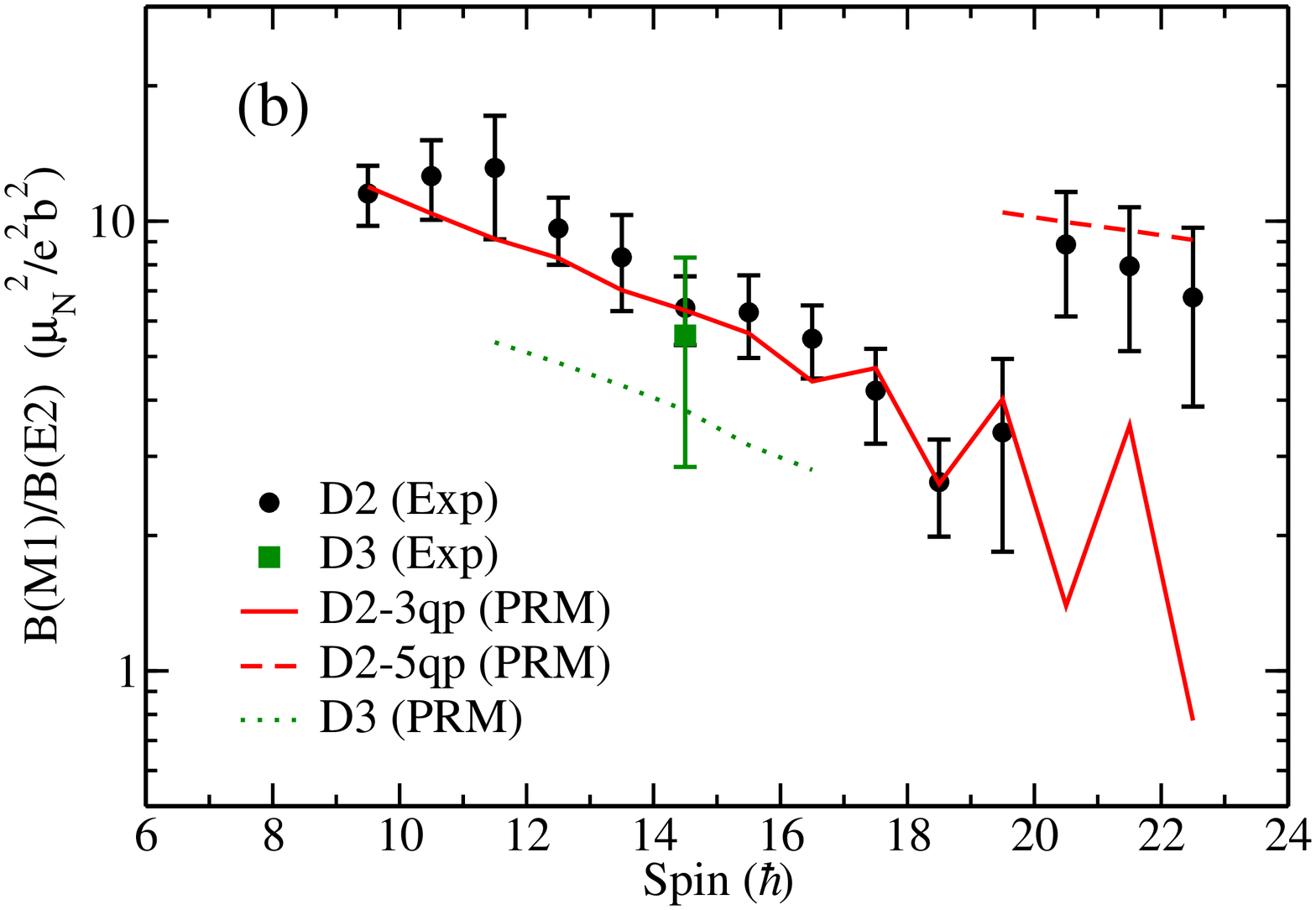}
 \vskip -0.5 cm
  \caption{(Color online) Comparison between experimental ratios of transitions probabilities B(M1)/B(E2) (symbols) and the particle-rotor calculations (lines) for the bands D1-D6.  }
  \label{fig5}
\end{figure}

A third quantity revealing the properties of the band is the ratio of reduced transition probabilities $B(M1)/B(E2)$, which are given together with the calculated values in Fig.~\ref{fig5}. One can observe the big difference between the $B(M1)/B(E2)$ values of the bands D1 and D2-D6 which are based on 1qp and 3qp (5qp) configurations, respectively.

In order to examine in detail the chiral character of the observed bands, the recently developed PRM has been used~\cite{135nd-prm,136nd-qibo}. The input deformation parameters $(\beta_2, \gamma)$ for PRM calculations are obtained from constrained CDFT calculations \cite{Meng2006Phys.Rev.C037303}. The moments of inertia are taken as irrotational flow type, i.e., $\mathcal{J}_k=\mathcal{J}_0\sin^2(\gamma-2k\pi/3)$, with $\mathcal{J}_0$ being adjusted to reproduce the trend of the energy spectra. In addition, for the electromagnetic transitions, the empirical intrinsic quadrupole moment $Q_0 = (3/\sqrt{5\pi})R_0^2Z\beta$ with $R_0 = 1.2A^{1/3}$ fm, and gyromagnetic ratios for rotor $g_R=Z/A$ and for nucleons $g_{p(n)} = g_l+(g_s-g_l)/(2l+1)$~[$g_l = 1(0)$ for protons (neutrons) and $g_s = 0.6g_{\textrm{free}}$]~\cite{ring2004nuclear} are adopted. 

The ground state band D1 with the configuration $\nu(1h_{11/2})^{-1}$ has been studied previously~\cite{135nd-zhu, 135nd-piel, 135nd-sd}. The deformation parameters obtained from the CDFT calculations are $\beta$ = 0.19 and $\gamma = 25.5^\circ$, which are similar to those used in Ref.~\cite{134nd-plunger}, in which the measured transition probabilities were well reproduced. The obtained energy spectra, which are shown in Fig.~\ref{fig3}, are in excellent agreement with the data in the low spin region with $\mathcal{J}_0=14.0~\hbar^2\textrm{MeV}^{-1}$. The deviation from the experimental data at high spin is due to the interaction with band D5 and to the variation of the moment of inertia which is not taken into account in the calculation. The corresponding calculated $B(M1)/B(E2)$ ratios, shown in Fig.~\ref{fig5} (a), are in very good agreement with the measured values.

The configurations  assigned to band D2-3qp and its continuation at high spins D2-5qp are $\pi[(1h_{11/2})^{1}(2d_{5/2})^{-1}]\otimes \nu(1h_{11/2})^{-1}$ and $\pi[(1h_{11/2})^{3}(2d_{5/2})^{-1}]\otimes \nu(1h_{11/2})^{-1}$, respectively. Their deformation parameters obtained from the CDFT calculations are $(\beta, \gamma)=(0.23, 22.5^\circ)$ and $(0.25, 15.6^\circ)$, respectively. To reproduce the $B(M1)/B(E2)$ for D2-3qp, a slightly smaller triaxial deformation $\gamma=17.0^\circ$ is used in the PRM calculations. In both calculations, a Coriolis attenuation factor $\xi=0.92$ is introduced. In addition, the used moments of inertia for D2-3qp and D2-5qp are $\mathcal{J}_0=25.0$ and $40.0~\hbar^2\textrm{MeV}^{-1}$, respectively. A larger $\mathcal{J}_0$ needed for D2-5qp is consistent with its lager single-particle alignment, as shown in Fig. \ref{fig4}. 
As one can see in Fig.~\ref{fig3}, the calculated excitation energies for band D2 are in excellent agreement with the experimental values  in the central and high spins ranges. At low spins, the calculated energies are lower than the experimental values by around 200 keV, which can be ascribed to the interaction with other low-lying levels. The calculated $B(M1)/B(E2)$ ratios of band D2 are compared with the experimental data in Fig.~\ref{fig5}~(b), in which we can see a good agreement for both D2-3qp and D2-5qp bands. 
 
The configuration assigned to the positive-parity doublet bands D3 and D4 is $\pi[(1h_{11/2})^{1}(1g_{7/2})^{-1}]\otimes \nu(1h_{11/2})^{-1}$, similar to that assigned to the corresponding bands 2 and 3 of the isotone nucleus $^{133}$Ce  \cite{Ayangeakaa2013Phys.Rev.Lett.172504}. In the PRM calculations, $(\beta=0.23, \gamma=21.0^\circ)$, $\mathcal{J}_0=28.0~\hbar^2\textrm{MeV}^{-1}$, and $\xi=0.94$ are employed. The calculated energy spectra presented in Fig.~\ref{fig3} reproduce well the experimental data. The energy separation between the bands  is nearly constant at $\approx$ 200 keV, reflecting similar moments of inertia, and supporting thus the chiral doublet bands interpretation. Due to very weak in-band crossover transitions, we could extract the $B(M1)/B(E2)$ ratio only for the level with spin $I = 29 \hbar$ of band D3, which is similar to that of band D2. The theoretical calculation is in agreement with the experimental values within the error bar [see Fig.~\ref{fig5}~(b)]. 
Based on this similarity between the observed bands in the two isotones $^{133}$Ce and $^{135}$Nd, we safely can interprete the bands D3 and D4 of $^{135}$Nd as chiral doublet bands based on the $\pi[(1h_{11/2})^{1}(1g_{7/2})^{-1}]\otimes \nu(1h_{11/2})^{-1}$ configuration.


The configuration adopted for the bands D5 and D6, which are among the best examples of chiral vibration, is $\pi(1h_{11/2})^{2}\otimes \nu(1h_{11/2})^{-1}$~\cite{135nd-zhu, chiral-vibration, 135nd-prm}. The deformation parameters from the CDFT are $(\beta, \gamma)=(0.24, 22.2^\circ)$, which are the same as those used in Ref. \cite{135nd-prm}. To reproduce the rapid increase of experimental $B(M1)/B(E2)$ value at $I=39/2\hbar$, a slightly smaller triaxial deformation $\gamma=20.2^\circ$ is used in the PRM calculations. In addition, $\mathcal{J}_0=23.5~\hbar^2\textrm{MeV}^{-1}$ and $\xi=0.98$ are adopted.  As one can see in Fig. \ref{fig3}, the calculated excitation energies of the two bands are in very good agreement with the experimental data. The measured transition probabilities $B(M1)$ and $B(E2)$ of the two bands were published in Ref.~\cite{chiral-vibration}. The $B(M1)/B(E2)$ ratios of the present work are very similar to those obtained from the results published in Ref. \cite{chiral-vibration}. The calculated and experimental $B(M1)/B(E2)$ ratios are showed in Fig. \ref{fig5} (a). One can see that the PRM values are in good agreement with the experimental data, including the sudden increase occurring at spin 39/2, which is induced by a sudden decrease of the $B(E2)$ values at spins above 39/2, and was interpreted as due to the transition from  chiral vibration to chiral rotation \cite{135nd-prm}. The success in reproducing the excitation energies and the electromagnetic transition probabilities of the bands D5 and D6 by the PRM calculations give a strong support to the configuration assignment of $\pi(1h_{11/2})^2\otimes \nu(1h_{11/2})^{-1}$.
  
\begin{figure}[htb]
  \centering
 \includegraphics[width=8.5cm]{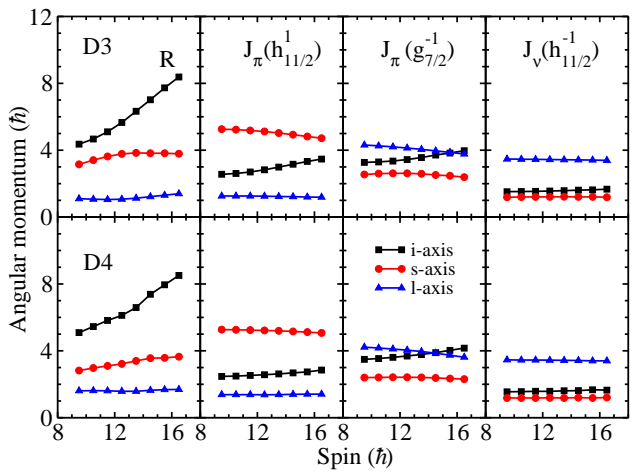}
  \caption{(Color online)  The root mean square components along the intermediate ($i-$, squares), short ($s-$, circles) and long ($l-$, triangles) axes of the rotor (R), valence protons (J$_{\pi}$), and valence neutrons (J$_{\nu}$) angular momenta calculated as functions of spin by PRM for the doublet bands D3 and D4 of $^{135}$Nd. }
  \label{ang-D3D4}
\end{figure}

In order to investigate to what extent the 3D chiral geometry is present in the two chiral doublets, we plotted the components of the angular momenta on the three axes (short, intermediate and long) of the intrinsic reference system. The results for bands D3 and D4 are shown in Fig. \ref{ang-D3D4}. Those for bands D5 and D6 are similar to those in Ref. \cite{135nd-prm}, hence here we do not present them once again. One can  observe a significative difference between the two chiral doublets: the positive-parity configuration assigned to bands D3 and D4 fulfills much better the chiral geometry, exhibiting equilibrated single-particle angular momenta along the three axes. The negative-parity configuration assigned to band D5 and D6 exhibits a higher single-particle angular momentum along the short axis, which brings the total angular momentum closer to the short-long principal plane and facilitates the chiral vibration at low spin.  

The observation of a second set of chiral doublet bands with positive-parity in addition to the previously known negative-parity chiral doublet, reveals the existence of the  M$\chi$D phenomenon in $^{135}$Nd. The presence of the M$\chi$D phenomenon in several nuclei of the $A \approx 130$ mass region ($^{133}$Ce, $^{135}$Nd, $^{136}$Nd), put in evidence the importance and solidity of the chiral symmetry in nuclei.

\section{Summary}
In summary, two sets of chiral doublet bands, (D3, D4) and (D5, D6), based on the 3qp configurations $\pi[(1h_{11/2})^{1}(1g_{7/2})^{-1}]\otimes \nu(1h_{11/2})^{-1}$ and $\pi(1h_{11/2})^{2}\otimes \nu(1h_{11/2})^{-1}$ have been identified in $^{135}$Nd.  The measured 
B(M1)/B(E2) ratios, the single-particle alignments and the near degeneracy of the bands strongly support the assignments of bands D4 and D6 as chiral partners of bands D3 and D5, respectively.
The observed doublet bands are compared with CDFT and PRM calculations, which nicely reproduce the experimental data. According to these results, the existence of the M$\chi$D phenomenon in nuclei of the $A\approx130$ mass region is strongly supported. The present results encourage us to continue the study of chirality in nuclei, both experimentally (measurement of lifetimes and search for chiral doublets in other nuclei) and theoretically. 

\section{Acknowledgements}
This work has been supported by the China Scholarship Council (CSC), CSC No. 201604910533; by the Academy of
Finland under the Finnish Centre of Excellence Programme (2012?2017); by the EU 7th Framework Programme Project
No. 262010 (ENSAR); by the National Research, Development and Innovation Fund of Hungary (Project 
  no. K128947), as well as by the European Regional Development Fund (Contract No. 
  GINOP-2.3.3-15-2016-00034); by the Polish National Science Centre (NCN) Grant No. 2013/10/M/ST2/00427; by
the Swedish Research Council under Grant No. 621-2014-5558; and by the National Natural Science Foundation of
China (Grants No. 11505242, No. 11305220, No. U1732139, No. 11775274, and No. 11575255). The use of germanium detectors from the GAMMAPOOL is acknowledged. The work of Q.B.C. is supported by Deutsche Forschungsgemeinschaft (DFG) and National Natural Science Foundation of China (NSFC) through funds provided to the Sino-German CRC 110 ``Symmetries and the Emergence of Structure in QCD" (DFG Grant No. TRR110
and NSFC Grant No. 11621131001). I.K. was supported by National Research, Development and Innovation Office - NKFIH, 
  contract number PD 124717. The authors are indebted to M. Loriggiola for his help in target preparation.

\bibliography{135nd}

\providecommand{\noopsort}[1]{}\providecommand{\singleletter}[1]{#1}%
\begin{thebibliography}{47}%
\makeatletter
\providecommand \@ifxundefined [1]{%
 \@ifx{#1\undefined}
}%
\providecommand \@ifnum [1]{%
 \ifnum #1\expandafter \@firstoftwo
 \else \expandafter \@secondoftwo
 \fi
}%
\providecommand \@ifx [1]{%
 \ifx #1\expandafter \@firstoftwo
 \else \expandafter \@secondoftwo
 \fi
}%
\providecommand \natexlab [1]{#1}%
\providecommand \enquote  [1]{``#1''}%
\providecommand \bibnamefont  [1]{#1}%
\providecommand \bibfnamefont [1]{#1}%
\providecommand \citenamefont [1]{#1}%
\providecommand \href@noop [0]{\@secondoftwo}%
\providecommand \href [0]{\begingroup \@sanitize@url \@href}%
\providecommand \@href[1]{\@@startlink{#1}\@@href}%
\providecommand \@@href[1]{\endgroup#1\@@endlink}%
\providecommand \@sanitize@url [0]{\catcode `\\12\catcode `\$12\catcode
  `\&12\catcode `\#12\catcode `\^12\catcode `\_12\catcode `\%12\relax}%
\providecommand \@@startlink[1]{}%
\providecommand \@@endlink[0]{}%
\providecommand \url  [0]{\begingroup\@sanitize@url \@url }%
\providecommand \@url [1]{\endgroup\@href {#1}{\urlprefix }}%
\providecommand \urlprefix  [0]{URL }%
\providecommand \Eprint [0]{\href }%
\providecommand \doibase [0]{http://dx.doi.org/}%
\providecommand \selectlanguage [0]{\@gobble}%
\providecommand \bibinfo  [0]{\@secondoftwo}%
\providecommand \bibfield  [0]{\@secondoftwo}%
\providecommand \translation [1]{[#1]}%
\providecommand \BibitemOpen [0]{}%
\providecommand \bibitemStop [0]{}%
\providecommand \bibitemNoStop [0]{.\EOS\space}%
\providecommand \EOS [0]{\spacefactor3000\relax}%
\providecommand \BibitemShut  [1]{\csname bibitem#1\endcsname}%
\let\auto@bib@innerbib\@empty
\bibitem [{\citenamefont {Frauendorf}\ and\ \citenamefont
  {Meng}(1997)}]{Stefan-Jie}%
  \BibitemOpen
  \bibfield  {author} {\bibinfo {author} {\bibfnamefont {S.}~\bibnamefont
  {Frauendorf}}\ and\ \bibinfo {author} {\bibfnamefont {J.}~\bibnamefont
  {Meng}},\ }\href {\doibase 10.1016/S0375-9474(97)00004-3} {\bibfield
  {journal} {\bibinfo  {journal} {Nucl. Phys. A}\ }\textbf {\bibinfo {volume}
  {617}},\ \bibinfo {pages} {131} (\bibinfo {year} {1997})}\BibitemShut
  {NoStop}%
\bibitem [{\citenamefont {Xiong}\ and\ \citenamefont
  {Wang}(2018)}]{adndt-chiral}%
  \BibitemOpen
  \bibfield  {author} {\bibinfo {author} {\bibfnamefont {B.~W.}\ \bibnamefont
  {Xiong}}\ and\ \bibinfo {author} {\bibfnamefont {Y.~Y.}\ \bibnamefont
  {Wang}},\ }\href {\doibase 10.1016/j.adt.2018.05.002} {\bibfield  {journal}
  {\bibinfo  {journal} {Atomic Data and Nuclear Data Tables}\ }\textbf
  {\bibinfo {volume} {111-112}},\ \bibinfo {pages} {193} (\bibinfo {year}
  {2018})}\BibitemShut {NoStop}%
\bibitem [{\citenamefont {Frauendorf}(2018)}]{frauendorf-2018}%
  \BibitemOpen
  \bibfield  {author} {\bibinfo {author} {\bibfnamefont {S.}~\bibnamefont
  {Frauendorf}},\ }\href {\doibase 10.1088/1402-4896/aaa2e9} {\bibfield
  {journal} {\bibinfo  {journal} {Int. J. Mod. Phys. E}\ }\textbf {\bibinfo
  {volume} {23}},\ \bibinfo {pages} {043003} (\bibinfo {year}
  {2018})}\BibitemShut {NoStop}%
\bibitem [{\citenamefont {Raduta}(2016)}]{raduta-2016}%
  \BibitemOpen
  \bibfield  {author} {\bibinfo {author} {\bibfnamefont {A.~A.}\ \bibnamefont
  {Raduta}},\ }\href {\doibase 10.1016//j.ppnp.2016.05.002} {\bibfield
  {journal} {\bibinfo  {journal} {Prog. Part. Nucl. Phys.}\ }\textbf {\bibinfo
  {volume} {90}},\ \bibinfo {pages} {241} (\bibinfo {year} {2016})}\BibitemShut
  {NoStop}%
\bibitem [{\citenamefont {Meng}\ and\ \citenamefont {Zhao}(2016)}]{meng-2016}%
  \BibitemOpen
  \bibfield  {author} {\bibinfo {author} {\bibfnamefont {J.}~\bibnamefont
  {Meng}}\ and\ \bibinfo {author} {\bibfnamefont {P.}~\bibnamefont {Zhao}},\
  }\href {\doibase 10.1088/0031-8949/91/053008} {\bibfield  {journal} {\bibinfo
   {journal} {Physics Scripta}\ }\textbf {\bibinfo {volume} {91}},\ \bibinfo
  {pages} {053008} (\bibinfo {year} {2016})}\BibitemShut {NoStop}%
\bibitem [{\citenamefont {Meng}\ \emph {et~al.}(2014)\citenamefont {Meng},
  \citenamefont {Chen},\ and\ \citenamefont {Zhang}}]{meng-2014}%
  \BibitemOpen
  \bibfield  {author} {\bibinfo {author} {\bibfnamefont {J.}~\bibnamefont
  {Meng}}, \bibinfo {author} {\bibfnamefont {Q.~B.}\ \bibnamefont {Chen}}, \
  and\ \bibinfo {author} {\bibfnamefont {S.~Q.}\ \bibnamefont {Zhang}},\ }\href
  {\doibase 10.11428/S0218301314300161} {\bibfield  {journal} {\bibinfo
  {journal} {Physics Scripta}\ }\textbf {\bibinfo {volume} {93}},\ \bibinfo
  {pages} {1430016} (\bibinfo {year} {2014})}\BibitemShut {NoStop}%
\bibitem [{\citenamefont {Meng}\ and\ \citenamefont
  {Zhang}(2010)}]{Meng2010JPhysG37}%
  \BibitemOpen
  \bibfield  {author} {\bibinfo {author} {\bibfnamefont {J.}~\bibnamefont
  {Meng}}\ and\ \bibinfo {author} {\bibfnamefont {S.~Q.}\ \bibnamefont
  {Zhang}},\ }\href {\doibase 10.1088/0954-3899/37/6/064025} {\bibfield
  {journal} {\bibinfo  {journal} {J. Phys. G.}\ }\textbf {\bibinfo {volume}
  {37}},\ \bibinfo {pages} {064025} (\bibinfo {year} {2010})}\BibitemShut
  {NoStop}%
\bibitem [{\citenamefont {Petrache~{\it et al.}}(2016)}]{133la-petrache}%
  \BibitemOpen
  \bibfield  {author} {\bibinfo {author} {\bibfnamefont {C.~M.}\ \bibnamefont
  {Petrache~{\it et al.}}},\ }\href {\doibase 10.1103/PhysRevC.94.064309}
  {\bibfield  {journal} {\bibinfo  {journal} {Phys. Rev. C}\ }\textbf {\bibinfo
  {volume} {94}},\ \bibinfo {pages} {064309} (\bibinfo {year}
  {2016})}\BibitemShut {NoStop}%
\bibitem [{\citenamefont {Ayangeakaa~{\it et
  al.}}(2013)}]{Ayangeakaa2013Phys.Rev.Lett.172504}%
  \BibitemOpen
  \bibfield  {author} {\bibinfo {author} {\bibfnamefont {A.~D.}\ \bibnamefont
  {Ayangeakaa~{\it et al.}}},\ }\href {\doibase 10.1103/PhysRevLett.110.172504}
  {\bibfield  {journal} {\bibinfo  {journal} {Phys. Rev. Lett.}\ }\textbf
  {\bibinfo {volume} {110}},\ \bibinfo {pages} {172504} (\bibinfo {year}
  {2013})}\BibitemShut {NoStop}%
\bibitem [{\citenamefont {Zhu~{\it et al.}}(2003)}]{135nd-zhu}%
  \BibitemOpen
  \bibfield  {author} {\bibinfo {author} {\bibfnamefont {S.}~\bibnamefont
  {Zhu~{\it et al.}}},\ }\href {\doibase 10.1103//PhysRevLett.97.132501}
  {\bibfield  {journal} {\bibinfo  {journal} {Phys. Rev. Lett.}\ }\textbf
  {\bibinfo {volume} {91}},\ \bibinfo {pages} {132501} (\bibinfo {year}
  {2003})}\BibitemShut {NoStop}%
\bibitem [{\citenamefont {Brant}\ and\ \citenamefont
  {Petrache}(2009)}]{137nd-brant}%
  \BibitemOpen
  \bibfield  {author} {\bibinfo {author} {\bibfnamefont {S.}~\bibnamefont
  {Brant}}\ and\ \bibinfo {author} {\bibfnamefont {C.~M.}\ \bibnamefont
  {Petrache}},\ }\href {\doibase 10.1103/PhysRevC.79.054326} {\bibfield
  {journal} {\bibinfo  {journal} {Phys. Rev. C}\ }\textbf {\bibinfo {volume}
  {79}},\ \bibinfo {pages} {054326} (\bibinfo {year} {2009})}\BibitemShut
  {NoStop}%
\bibitem [{\citenamefont {Kuti~{\it et
  al.}}(2014)}]{Kuti2014Phys.Rev.Lett.032501}%
  \BibitemOpen
  \bibfield  {author} {\bibinfo {author} {\bibfnamefont {I.}~\bibnamefont
  {Kuti~{\it et al.}}},\ }\href {\doibase 10.1103/PhysRevLett.113.032501}
  {\bibfield  {journal} {\bibinfo  {journal} {Phys. Rev. Lett.}\ }\textbf
  {\bibinfo {volume} {113}},\ \bibinfo {pages} {032501} (\bibinfo {year}
  {2014})}\BibitemShut {NoStop}%
\bibitem [{\citenamefont {Alc\'antara-N\'u\~nez{\it et
  al.}}(2004)}]{Alcantara2004}%
  \BibitemOpen
  \bibfield  {author} {\bibinfo {author} {\bibfnamefont {J.~A.}\ \bibnamefont
  {Alc\'antara-N\'u\~nez{\it et al.}}},\ }\href {\doibase
  10.1103/PhysRevC.69.024317} {\bibfield  {journal} {\bibinfo  {journal} {Phys.
  Rev. C}\ }\textbf {\bibinfo {volume} {69}},\ \bibinfo {pages} {024317}
  (\bibinfo {year} {2004})}\BibitemShut {NoStop}%
\bibitem [{\citenamefont {Timar~{\it et al}.}(2004)}]{105Rh}%
  \BibitemOpen
  \bibfield  {author} {\bibinfo {author} {\bibfnamefont {J.}~\bibnamefont
  {Timar~{\it et al}.}},\ }\href {\doibase 10.1016j.physletb.2004.07.050}
  {\bibfield  {journal} {\bibinfo  {journal} {Phys. Lett. B}\ }\textbf
  {\bibinfo {volume} {598}},\ \bibinfo {pages} {178} (\bibinfo {year}
  {2004})}\BibitemShut {NoStop}%
\bibitem [{\citenamefont {Luo~{\it et al}.}(2004)}]{111-113Rh-luo}%
  \BibitemOpen
  \bibfield  {author} {\bibinfo {author} {\bibfnamefont {Y.~X.}\ \bibnamefont
  {Luo~{\it et al}.}},\ }\href {\doibase 10.1103/PhysRevC.69.024315} {\bibfield
   {journal} {\bibinfo  {journal} {Phys. Rev. C}\ }\textbf {\bibinfo {volume}
  {69}},\ \bibinfo {pages} {024315} (\bibinfo {year} {2004})}\BibitemShut
  {NoStop}%
\bibitem [{\citenamefont {Timar~{\it et al}.}(2007)}]{105Ag}%
  \BibitemOpen
  \bibfield  {author} {\bibinfo {author} {\bibfnamefont {J.}~\bibnamefont
  {Timar~{\it et al}.}},\ }\href {\doibase 10.1103/PhysRevC.76.024307}
  {\bibfield  {journal} {\bibinfo  {journal} {Phys. Rev. C}\ }\textbf {\bibinfo
  {volume} {76}},\ \bibinfo {pages} {024307} (\bibinfo {year}
  {2007})}\BibitemShut {NoStop}%
\bibitem [{\citenamefont {He~{\it et al}.}(2004)}]{107Ag}%
  \BibitemOpen
  \bibfield  {author} {\bibinfo {author} {\bibfnamefont {C.~Y.}\ \bibnamefont
  {He~{\it et al}.}},\ }\href {\doibase 10.1088/1009-0630/14/6/18} {\bibfield
  {journal} {\bibinfo  {journal} {Plasma Sci. Technol}\ }\textbf {\bibinfo
  {volume} {14}},\ \bibinfo {pages} {518} (\bibinfo {year} {2004})}\BibitemShut
  {NoStop}%
\bibitem [{\citenamefont {Petrache~{\it et al.}}(2018)}]{136nd-PRCR}%
  \BibitemOpen
  \bibfield  {author} {\bibinfo {author} {\bibfnamefont {C.~M.}\ \bibnamefont
  {Petrache~{\it et al.}}},\ }\href {\doibase 10.1103/PhysRevC.97.041304}
  {\bibfield  {journal} {\bibinfo  {journal} {Phys. Rev. C}\ }\textbf {\bibinfo
  {volume} {97}},\ \bibinfo {pages} {041304(R)} (\bibinfo {year}
  {2018})}\BibitemShut {NoStop}%
\bibitem [{\citenamefont {Petrache}\ \emph {et~al.}(2012)\citenamefont
  {Petrache}, \citenamefont {Frauendorf}, \citenamefont {Matsuzaki},
  \citenamefont {Leguillon}, \citenamefont {Zerrouki}, \citenamefont {Lunardi},
  \citenamefont {Bazzacco}, \citenamefont {Ur}, \citenamefont {Farnea},
  \citenamefont {Rossi~Alvarez}, \citenamefont {Venturelli},\ and\
  \citenamefont {de~Angelis}}]{138nd-tilted}%
  \BibitemOpen
  \bibfield  {author} {\bibinfo {author} {\bibfnamefont {C.~M.}\ \bibnamefont
  {Petrache}}, \bibinfo {author} {\bibfnamefont {S.}~\bibnamefont
  {Frauendorf}}, \bibinfo {author} {\bibfnamefont {M.}~\bibnamefont
  {Matsuzaki}}, \bibinfo {author} {\bibfnamefont {R.}~\bibnamefont
  {Leguillon}}, \bibinfo {author} {\bibfnamefont {T.}~\bibnamefont {Zerrouki}},
  \bibinfo {author} {\bibfnamefont {S.}~\bibnamefont {Lunardi}}, \bibinfo
  {author} {\bibfnamefont {D.}~\bibnamefont {Bazzacco}}, \bibinfo {author}
  {\bibfnamefont {C.~A.}\ \bibnamefont {Ur}}, \bibinfo {author} {\bibfnamefont
  {E.}~\bibnamefont {Farnea}}, \bibinfo {author} {\bibfnamefont
  {C.}~\bibnamefont {Rossi~Alvarez}}, \bibinfo {author} {\bibfnamefont
  {R.}~\bibnamefont {Venturelli}}, \ and\ \bibinfo {author} {\bibfnamefont
  {G.}~\bibnamefont {de~Angelis}},\ }\href {\doibase
  10.1103/PhysRevC.86.044321} {\bibfield  {journal} {\bibinfo  {journal} {Phys.
  Rev. C}\ }\textbf {\bibinfo {volume} {86}},\ \bibinfo {pages} {044321}
  (\bibinfo {year} {2012})}\BibitemShut {NoStop}%
\bibitem [{\citenamefont {Liu~{\it et
  al.}}(2016)}]{Liu2016Phys.Rev.Lett.112501}%
  \BibitemOpen
  \bibfield  {author} {\bibinfo {author} {\bibfnamefont {C.}~\bibnamefont
  {Liu~{\it et al.}}},\ }\href {\doibase 10.1103/PhysRevLett.116.112501}
  {\bibfield  {journal} {\bibinfo  {journal} {Phys. Rev. Lett.}\ }\textbf
  {\bibinfo {volume} {116}},\ \bibinfo {pages} {112501} (\bibinfo {year}
  {2016})}\BibitemShut {NoStop}%
\bibitem [{\citenamefont {Roy~{\it et al.}}(2018)}]{195Tl-roy}%
  \BibitemOpen
  \bibfield  {author} {\bibinfo {author} {\bibfnamefont {T.}~\bibnamefont
  {Roy~{\it et al.}}},\ }\href {\doibase 10.1016j.physletb.2018.06.033}
  {\bibfield  {journal} {\bibinfo  {journal} {Phys. Lett. B}\ }\textbf
  {\bibinfo {volume} {782}},\ \bibinfo {pages} {768} (\bibinfo {year}
  {2018})}\BibitemShut {NoStop}%
\bibitem [{\citenamefont {Qi}\ \emph {et~al.}(2013)\citenamefont {Qi},
  \citenamefont {Jia}, \citenamefont {Zhang}, \citenamefont {Liu},\ and\
  \citenamefont {Wang}}]{107ag-qi}%
  \BibitemOpen
  \bibfield  {author} {\bibinfo {author} {\bibfnamefont {B.}~\bibnamefont
  {Qi}}, \bibinfo {author} {\bibfnamefont {H.}~\bibnamefont {Jia}}, \bibinfo
  {author} {\bibfnamefont {N.~B.}\ \bibnamefont {Zhang}}, \bibinfo {author}
  {\bibfnamefont {C.}~\bibnamefont {Liu}}, \ and\ \bibinfo {author}
  {\bibfnamefont {S.~Y.}\ \bibnamefont {Wang}},\ }\href {\doibase
  10.1103/PhysRevC.88.027302} {\bibfield  {journal} {\bibinfo  {journal} {Phys.
  Rev. C}\ }\textbf {\bibinfo {volume} {88}},\ \bibinfo {pages} {027302}
  (\bibinfo {year} {2013})}\BibitemShut {NoStop}%
\bibitem [{\citenamefont {Meng}\ \emph {et~al.}(2006)\citenamefont {Meng},
  \citenamefont {Peng}, \citenamefont {Zhang},\ and\ \citenamefont
  {Zhou}}]{Meng2006Phys.Rev.C037303}%
  \BibitemOpen
  \bibfield  {author} {\bibinfo {author} {\bibfnamefont {J.}~\bibnamefont
  {Meng}}, \bibinfo {author} {\bibfnamefont {J.}~\bibnamefont {Peng}}, \bibinfo
  {author} {\bibfnamefont {S.~Q.}\ \bibnamefont {Zhang}}, \ and\ \bibinfo
  {author} {\bibfnamefont {S.-G.}\ \bibnamefont {Zhou}},\ }\href {\doibase
  10.1103/PhysRevC.73.037303} {\bibfield  {journal} {\bibinfo  {journal} {Phys.
  Rev. C}\ }\textbf {\bibinfo {volume} {73}},\ \bibinfo {pages} {037303}
  (\bibinfo {year} {2006})}\BibitemShut {NoStop}%
\bibitem [{\citenamefont {Peng}\ \emph {et~al.}(2008)\citenamefont {Peng},
  \citenamefont {Sagawa}, \citenamefont {Zhang}, \citenamefont {Yao},
  \citenamefont {Zhang},\ and\ \citenamefont
  {Meng}}]{Peng2008Phys.Rev.C024309}%
  \BibitemOpen
  \bibfield  {author} {\bibinfo {author} {\bibfnamefont {J.}~\bibnamefont
  {Peng}}, \bibinfo {author} {\bibfnamefont {H.}~\bibnamefont {Sagawa}},
  \bibinfo {author} {\bibfnamefont {S.~Q.}\ \bibnamefont {Zhang}}, \bibinfo
  {author} {\bibfnamefont {J.~M.}\ \bibnamefont {Yao}}, \bibinfo {author}
  {\bibfnamefont {Y.}~\bibnamefont {Zhang}}, \ and\ \bibinfo {author}
  {\bibfnamefont {J.}~\bibnamefont {Meng}},\ }\href {\doibase
  10.1103/PhysRevC.77.024309} {\bibfield  {journal} {\bibinfo  {journal} {Phys.
  Rev. C}\ }\textbf {\bibinfo {volume} {77}},\ \bibinfo {pages} {024309}
  (\bibinfo {year} {2008})}\BibitemShut {NoStop}%
\bibitem [{\citenamefont {Yao}\ \emph {et~al.}(2009)\citenamefont {Yao},
  \citenamefont {Qi}, \citenamefont {Zhang}, \citenamefont {Peng},
  \citenamefont {Wang},\ and\ \citenamefont {Meng}}]{Yao2009Phys.Rev.C067302}%
  \BibitemOpen
  \bibfield  {author} {\bibinfo {author} {\bibfnamefont {J.~M.}\ \bibnamefont
  {Yao}}, \bibinfo {author} {\bibfnamefont {B.}~\bibnamefont {Qi}}, \bibinfo
  {author} {\bibfnamefont {S.~Q.}\ \bibnamefont {Zhang}}, \bibinfo {author}
  {\bibfnamefont {J.}~\bibnamefont {Peng}}, \bibinfo {author} {\bibfnamefont
  {S.~Y.}\ \bibnamefont {Wang}}, \ and\ \bibinfo {author} {\bibfnamefont
  {J.}~\bibnamefont {Meng}},\ }\href {\doibase 10.1103/PhysRevC.79.067302}
  {\bibfield  {journal} {\bibinfo  {journal} {Phys. Rev. C}\ }\textbf {\bibinfo
  {volume} {79}},\ \bibinfo {pages} {067302} (\bibinfo {year}
  {2009})}\BibitemShut {NoStop}%
\bibitem [{\citenamefont {Li}\ \emph {et~al.}(2011)\citenamefont {Li},
  \citenamefont {Zhang},\ and\ \citenamefont {Meng}}]{Li2011Phys.Rev.C037301}%
  \BibitemOpen
  \bibfield  {author} {\bibinfo {author} {\bibfnamefont {J.}~\bibnamefont
  {Li}}, \bibinfo {author} {\bibfnamefont {S.~Q.}\ \bibnamefont {Zhang}}, \
  and\ \bibinfo {author} {\bibfnamefont {J.}~\bibnamefont {Meng}},\ }\href
  {\doibase 10.1103/PhysRevC.83.037301} {\bibfield  {journal} {\bibinfo
  {journal} {Phys. Rev. C}\ }\textbf {\bibinfo {volume} {83}},\ \bibinfo
  {pages} {037301} (\bibinfo {year} {2011})}\BibitemShut {NoStop}%
\bibitem [{\citenamefont {Droste}\ \emph {et~al.}(2009)\citenamefont {Droste},
  \citenamefont {Rohozinski}, \citenamefont {Starosta},\ and\ \citenamefont
  {Prochniak}}]{Droste2009EPJA}%
  \BibitemOpen
  \bibfield  {author} {\bibinfo {author} {\bibfnamefont {C.}~\bibnamefont
  {Droste}}, \bibinfo {author} {\bibfnamefont {S.}~\bibnamefont {Rohozinski}},
  \bibinfo {author} {\bibfnamefont {K.}~\bibnamefont {Starosta}}, \ and\
  \bibinfo {author} {\bibfnamefont {E.}~\bibnamefont {Prochniak}, \bibfnamefont
  {L.~andGrodner}},\ }\href {\doibase 10.1140/epja/i2009-10860-0} {\bibfield
  {journal} {\bibinfo  {journal} {Eur. Phys. J. A}\ }\textbf {\bibinfo {volume}
  {42}},\ \bibinfo {pages} {79} (\bibinfo {year} {2009})}\BibitemShut {NoStop}%
\bibitem [{\citenamefont {Chen}\ \emph {et~al.}(2010)\citenamefont {Chen},
  \citenamefont {Yao}, \citenamefont {Zhang},\ and\ \citenamefont
  {Qi}}]{Chen2010PRC}%
  \BibitemOpen
  \bibfield  {author} {\bibinfo {author} {\bibfnamefont {Q.~B.}\ \bibnamefont
  {Chen}}, \bibinfo {author} {\bibfnamefont {J.~M.}\ \bibnamefont {Yao}},
  \bibinfo {author} {\bibfnamefont {S.~Q.}\ \bibnamefont {Zhang}}, \ and\
  \bibinfo {author} {\bibfnamefont {B.}~\bibnamefont {Qi}},\ }\href {\doibase
  10.1103/PhysRevC.82.067302} {\bibfield  {journal} {\bibinfo  {journal} {Phys.
  Rev. C}\ }\textbf {\bibinfo {volume} {82}},\ \bibinfo {pages} {067302}
  (\bibinfo {year} {2010})}\BibitemShut {NoStop}%
\bibitem [{\citenamefont {Hamamoto}(2013)}]{Hamamoto2013PRC}%
  \BibitemOpen
  \bibfield  {author} {\bibinfo {author} {\bibfnamefont {I.}~\bibnamefont
  {Hamamoto}},\ }\href {\doibase 10.1103/PhysRevC.88.024327} {\bibfield
  {journal} {\bibinfo  {journal} {Phys. Rev. C}\ }\textbf {\bibinfo {volume}
  {88}},\ \bibinfo {pages} {024327} (\bibinfo {year} {2013})}\BibitemShut
  {NoStop}%
\bibitem [{\citenamefont {Zhang}\ and\ \citenamefont
  {CHen}(2016)}]{Zhang2016CPC}%
  \BibitemOpen
  \bibfield  {author} {\bibinfo {author} {\bibfnamefont {H.}~\bibnamefont
  {Zhang}}\ and\ \bibinfo {author} {\bibfnamefont {Q.~B.}\ \bibnamefont
  {CHen}},\ }\href {\doibase 10.1088/1674-1137/40/2/024102} {\bibfield
  {journal} {\bibinfo  {journal} {Chin. Phys. C}\ }\textbf {\bibinfo {volume}
  {40}},\ \bibinfo {pages} {024101} (\bibinfo {year} {2016})}\BibitemShut
  {NoStop}%
\bibitem [{\citenamefont {Chen}\ \emph {et~al.}(2018)\citenamefont {Chen},
  \citenamefont {Lv}, \citenamefont {Petrache},\ and\ \citenamefont
  {Meng}}]{136nd-qibo}%
  \BibitemOpen
  \bibfield  {author} {\bibinfo {author} {\bibfnamefont {Q.~B.}\ \bibnamefont
  {Chen}}, \bibinfo {author} {\bibfnamefont {B.~F.}\ \bibnamefont {Lv}},
  \bibinfo {author} {\bibfnamefont {C.~M.}\ \bibnamefont {Petrache}}, \ and\
  \bibinfo {author} {\bibfnamefont {J.}~\bibnamefont {Meng}},\ }\href {\doibase
  https://doi.org/10.1016/j.physletb.2018.06.030} {\bibfield  {journal}
  {\bibinfo  {journal} {Phys. Lett. B}\ }\textbf {\bibinfo {volume} {782}},\
  \bibinfo {pages} {744} (\bibinfo {year} {2018})}\BibitemShut {NoStop}%
\bibitem [{\citenamefont {Raduta}\ \emph {et~al.}(2016)\citenamefont {Raduta},
  \citenamefont {Raduta},\ and\ \citenamefont {Petrache}}]{raduta-jpg-2016}%
  \BibitemOpen
  \bibfield  {author} {\bibinfo {author} {\bibfnamefont {A.~A.}\ \bibnamefont
  {Raduta}}, \bibinfo {author} {\bibfnamefont {A.~H.}\ \bibnamefont {Raduta}},
  \ and\ \bibinfo {author} {\bibfnamefont {C.~M.}\ \bibnamefont {Petrache}},\
  }\href {\doibase 10.1088/0954-3899/43/9/095107} {\bibfield  {journal}
  {\bibinfo  {journal} {J. Phys. G.}\ }\textbf {\bibinfo {volume} {43}},\
  \bibinfo {pages} {095107} (\bibinfo {year} {2016})}\BibitemShut {NoStop}%
\bibitem [{\citenamefont {Raduta}\ \emph {et~al.}(2017)\citenamefont {Raduta},
  \citenamefont {Raduta},\ and\ \citenamefont {Raduta}}]{raduta-jpg-2017}%
  \BibitemOpen
  \bibfield  {author} {\bibinfo {author} {\bibfnamefont {A.~A.}\ \bibnamefont
  {Raduta}}, \bibinfo {author} {\bibfnamefont {C.~M.}\ \bibnamefont {Raduta}},
  \ and\ \bibinfo {author} {\bibfnamefont {A.~H.}\ \bibnamefont {Raduta}},\
  }\href {\doibase 10.1088/1361-6471/aa5af1} {\bibfield  {journal} {\bibinfo
  {journal} {J. Phys. G.}\ }\textbf {\bibinfo {volume} {44}},\ \bibinfo {pages}
  {045102} (\bibinfo {year} {2017})}\BibitemShut {NoStop}%
\bibitem [{\citenamefont {Qi}\ \emph {et~al.}(2009)\citenamefont {Qi},
  \citenamefont {Zhang}, \citenamefont {Meng}, \citenamefont {Wang},\ and\
  \citenamefont {Frauendorf}}]{135nd-prm}%
  \BibitemOpen
  \bibfield  {author} {\bibinfo {author} {\bibfnamefont {B.}~\bibnamefont
  {Qi}}, \bibinfo {author} {\bibfnamefont {S.~Q.}\ \bibnamefont {Zhang}},
  \bibinfo {author} {\bibfnamefont {J.}~\bibnamefont {Meng}}, \bibinfo {author}
  {\bibfnamefont {S.~Y.}\ \bibnamefont {Wang}}, \ and\ \bibinfo {author}
  {\bibfnamefont {S.}~\bibnamefont {Frauendorf}},\ }\href {\doibase
  10.1016j.physletb.2009.02.061} {\bibfield  {journal} {\bibinfo  {journal}
  {Phys. Lett. B}\ }\textbf {\bibinfo {volume} {675}},\ \bibinfo {pages} {175}
  (\bibinfo {year} {2009})}\BibitemShut {NoStop}%
\bibitem [{\citenamefont {Zhu}()}]{135nd-thesis}%
  \BibitemOpen
  \bibfield  {author} {\bibinfo {author} {\bibfnamefont {S.}~\bibnamefont
  {Zhu}},\ }\href@noop {} {\bibinfo  {journal} {{\it PhD thesis, 2004}}\
  }\BibitemShut {NoStop}%
\bibitem [{\citenamefont {Mukhopadhyay~{\it et al.}}(2007)}]{chiral-vibration}%
  \BibitemOpen
\bibfield  {journal} {  }\bibfield  {author} {\bibinfo {author} {\bibfnamefont
  {S.}~\bibnamefont {Mukhopadhyay~{\it et al.}}},\ }\href {\doibase
  10.1103/PhysRevLett.99.172501} {\bibfield  {journal} {\bibinfo  {journal}
  {Phys. Rev. Lett.}\ }\textbf {\bibinfo {volume} {99}},\ \bibinfo {pages}
  {172501} (\bibinfo {year} {2007})}\BibitemShut {NoStop}%
\bibitem [{\citenamefont {Zhao}\ \emph {et~al.}(2015)\citenamefont {Zhao},
  \citenamefont {Zhang},\ and\ \citenamefont {Meng}}]{135nd-pairing}%
  \BibitemOpen
  \bibfield  {author} {\bibinfo {author} {\bibfnamefont {P.~W.}\ \bibnamefont
  {Zhao}}, \bibinfo {author} {\bibfnamefont {S.~Q.}\ \bibnamefont {Zhang}}, \
  and\ \bibinfo {author} {\bibfnamefont {J.}~\bibnamefont {Meng}},\ }\href
  {\doibase 10.1103/PhysRevC.92.034319} {\bibfield  {journal} {\bibinfo
  {journal} {Phys. Rev. C}\ }\textbf {\bibinfo {volume} {92}},\ \bibinfo
  {pages} {034319} (\bibinfo {year} {2015})}\BibitemShut {NoStop}%
\bibitem [{\citenamefont {Lv~{\it et al.}}(2018)}]{lv}%
  \BibitemOpen
  \bibfield  {author} {\bibinfo {author} {\bibfnamefont {B.~F.}\ \bibnamefont
  {Lv~{\it et al.}}},\ }\href {\doibase 10.1103/PhysRevC.98.044304} {\bibfield
  {journal} {\bibinfo  {journal} {Phys. Rev. C}\ }\textbf {\bibinfo {volume}
  {98}},\ \bibinfo {pages} {044304} (\bibinfo {year} {2018})}\BibitemShut
  {NoStop}%
\bibitem [{\citenamefont {Kr\"{a}mer-Flecken}\ \emph
  {et~al.}(1989)\citenamefont {Kr\"{a}mer-Flecken}, \citenamefont {Morek},
  \citenamefont {Lieder}, \citenamefont {Gast}, \citenamefont {Hebbinghaus},
  \citenamefont {J\"{a}ger},\ and\ \citenamefont
  {Urban}}]{KramerFlecken1989333}%
  \BibitemOpen
  \bibfield  {author} {\bibinfo {author} {\bibfnamefont {A.}~\bibnamefont
  {Kr\"{a}mer-Flecken}}, \bibinfo {author} {\bibfnamefont {T.}~\bibnamefont
  {Morek}}, \bibinfo {author} {\bibfnamefont {R.~M.}\ \bibnamefont {Lieder}},
  \bibinfo {author} {\bibfnamefont {W.}~\bibnamefont {Gast}}, \bibinfo {author}
  {\bibfnamefont {G.}~\bibnamefont {Hebbinghaus}}, \bibinfo {author}
  {\bibfnamefont {H.~M.}\ \bibnamefont {J\"{a}ger}}, \ and\ \bibinfo {author}
  {\bibfnamefont {W.}~\bibnamefont {Urban}},\ }\href {\doibase
  10.1016/0168-9002(89)90706-7} {\bibfield  {journal} {\bibinfo  {journal}
  {Nucl. Instrum. Meth. Phys. Res. A}\ }\textbf {\bibinfo {volume} {275}},\
  \bibinfo {pages} {333 } (\bibinfo {year} {1989})}\BibitemShut {NoStop}%
\bibitem [{\citenamefont {Chiara~{\it et al.}}(2007)}]{Chiara.75.054305}%
  \BibitemOpen
  \bibfield  {author} {\bibinfo {author} {\bibfnamefont {C.~J.}\ \bibnamefont
  {Chiara~{\it et al.}}},\ }\href {\doibase 10.1103/PhysRevC.75.054305}
  {\bibfield  {journal} {\bibinfo  {journal} {Phys. Rev. C}\ }\textbf {\bibinfo
  {volume} {75}},\ \bibinfo {pages} {054305} (\bibinfo {year}
  {2007})}\BibitemShut {NoStop}%
\bibitem [{\citenamefont {Herz\'a\ifmmode\check{n}\else\v{n}\fi{}~{\it et
  al.}}(2015)}]{PhysRevC.92.044310}%
  \BibitemOpen
  \bibfield  {author} {\bibinfo {author} {\bibfnamefont {A.}~\bibnamefont
  {Herz\'a\ifmmode\check{n}\else\v{n}\fi{}~{\it et al.}}},\ }\href {\doibase
  10.1103/PhysRevC.92.044310} {\bibfield  {journal} {\bibinfo  {journal} {Phys.
  Rev. C}\ }\textbf {\bibinfo {volume} {92}},\ \bibinfo {pages} {044310}
  (\bibinfo {year} {2015})}\BibitemShut {NoStop}%
\bibitem [{\citenamefont {Starosta~{\it et al.}}(1999)}]{STAROSTA199916}%
  \BibitemOpen
  \bibfield  {author} {\bibinfo {author} {\bibfnamefont {K.}~\bibnamefont
  {Starosta~{\it et al.}}},\ }\href {\doibase
  https://doi.org/10.1016/S0168-9002(98)01220-0} {\bibfield  {journal}
  {\bibinfo  {journal} {Nucl. Instrum. Meth. Phys. Res. A}\ }\textbf {\bibinfo
  {volume} {423}},\ \bibinfo {pages} {16} (\bibinfo {year} {1999})}\BibitemShut
  {NoStop}%
\bibitem [{\citenamefont {Piel}\ \emph {et~al.}(1987)\citenamefont {Piel},
  \citenamefont {Beausang}, \citenamefont {Fossan}, \citenamefont
  {Hildingsson},\ and\ \citenamefont {Paul}}]{135nd-piel}%
  \BibitemOpen
  \bibfield  {author} {\bibinfo {author} {\bibfnamefont {W.~F.}\ \bibnamefont
  {Piel}}, \bibinfo {author} {\bibfnamefont {C.~W.}\ \bibnamefont {Beausang}},
  \bibinfo {author} {\bibfnamefont {D.~B.}\ \bibnamefont {Fossan}}, \bibinfo
  {author} {\bibfnamefont {L.}~\bibnamefont {Hildingsson}}, \ and\ \bibinfo
  {author} {\bibfnamefont {E.~S.}\ \bibnamefont {Paul}},\ }\href {\doibase
  10.1103/PhysRevC.35.959} {\bibfield  {journal} {\bibinfo  {journal} {Phys.
  Rev. C}\ }\textbf {\bibinfo {volume} {35}},\ \bibinfo {pages} {959} (\bibinfo
  {year} {1987})}\BibitemShut {NoStop}%
\bibitem [{\citenamefont {Petrache{\it et al}.}()}]{135nd-wobb}%
  \BibitemOpen
  \bibfield  {author} {\bibinfo {author} {\bibfnamefont {C.~M.}\ \bibnamefont
  {Petrache{\it et al}.}},\ }\href@noop {} {\bibinfo  {journal} {{\it to be
  published}}\ }\BibitemShut {NoStop}%
\bibitem [{\citenamefont {Ring}\ and\ \citenamefont
  {Schuck}(2004)}]{ring2004nuclear}%
  \BibitemOpen
\bibfield  {journal} {  }\bibfield  {author} {\bibinfo {author} {\bibfnamefont
  {P.}~\bibnamefont {Ring}}\ and\ \bibinfo {author} {\bibfnamefont
  {P.}~\bibnamefont {Schuck}},\ }\href@noop {} {}\ (\bibinfo  {publisher}
  {Springer Science \& Business Media},\ \bibinfo {year} {2004})\BibitemShut
  {NoStop}%
\bibitem [{\citenamefont {Beck~{\it et al.}}(1987)}]{135nd-sd}%
  \BibitemOpen
  \bibfield  {author} {\bibinfo {author} {\bibfnamefont {E.~M.}\ \bibnamefont
  {Beck~{\it et al.}}},\ }\href {\doibase 10.1103/PhysRevLett.58.2182}
  {\bibfield  {journal} {\bibinfo  {journal} {Phys. Rev. Lett.}\ }\textbf
  {\bibinfo {volume} {58}},\ \bibinfo {pages} {2182} (\bibinfo {year}
  {1987})}\BibitemShut {NoStop}%
\bibitem [{\citenamefont {Klemme~{\it et al}.}(1999)}]{134nd-plunger}%
  \BibitemOpen
  \bibfield  {author} {\bibinfo {author} {\bibfnamefont {T.}~\bibnamefont
  {Klemme~{\it et al}.}},\ }\href {\doibase 10.1103/PhysRevC.60.034301}
  {\bibfield  {journal} {\bibinfo  {journal} {Phys. Rev. C}\ }\textbf {\bibinfo
  {volume} {60}},\ \bibinfo {pages} {034301} (\bibinfo {year}
  {1999})}\BibitemShut {NoStop}%
\end{thebibliography}%
\newpage

\begin{longtable*}{ccccccc}
\caption{\label{135nd-table} Experimental information including the $\gamma$-ray energies, energies of the initial levels $E_{i}$, intensities $I_\gamma$, anisotropies $R_{DCO}$ and or $R_{ac}$, multipolarities, and the spin-parity assignments to the observed states in $^{135}$Nd. The transitions listed with increasing energy are grouped in bands. The deduced values for $R_{DCO}$ with a stretched quadrupole gate are $\approx$ 1 for stretched quadrupole and $\approx$ 0.46 for dipole transitions, while the ratio is close to 1 for a dipole and 2.1 for a quadrupole transition when the gate is set on a dipole transition. The $R_{ac}$  values for  stretched dipole and quadrupole transitions are  $\approx$ 0.8 and $\approx$ 1.4. }\\
\hhline{=======}
$E_\gamma$$^a$  & E$_i$ (keV)   & $I_\gamma$$^b$ &$R_{DCO}$$^c$ &$R_{ac}$$^d$  & Multipolarity  & $J^{\pi}_i$ $\rightarrow$ $J^{\pi}_f$\\
\hline
\endfirsthead
\multicolumn{7}{c}%
{\tablename\ \thetable.  (\textit{Continued.})} \\
\hhline{=======}
$E_\gamma$$^a$  & E$_i$ (keV)   & $I_\gamma$$^b$ &$R_{DCO}$$^c$ &$R_{ac}$$^d$  & Multipolarity  & $J^{\pi}_i$ $\rightarrow$ $J^{\pi}_f$\\
\hline
\endhead
\hline \multicolumn{7}{r}{\textit{}} \\
\endfoot
\hline
\endlastfoot
  $\bf Band~D1$ & & &         &            & &                                               \\
          198.8             & 198.8       & 100.0               &0.49(5)$^e$        &               & M1            & 11/2$^-$ $\rightarrow$ ~9/2$^-$  \\
          232.6             & 793.1       & 12.5(19)            &0.42(7)$^e$       &               & M1            & 15/2$^{-}$ $\rightarrow$ 13/2$^{-}$ \\
          250.6             & 1520.4     & 2.9(5)              &0.47(8)$^e$         &               & M1           & 19/2$^-$ $\rightarrow$ 17/2$^-$    \\
          270.1             & 2375.6     & 1.2(3)              &0.24(5)$^e$         &               & M1/E2      & 23/2$^-$ $\rightarrow$ 21/2$^-$    \\    
          276.2             & 3319.5     & 0.10(5)            &                            &               &                  & 27/2$^-$ $\rightarrow$  25/2$^-$ \\
          361.7             & 560.5       & 31(3)               &0.43(6)$^e$         &               & M1           & 13/2$^-$ $\rightarrow$ 11/2$^-$    \\
          476.7             & 1269.8     &14.0(25)           &0.40(8)$^e$         &               & M1           & 17/2$^-$ $\rightarrow$ 15/2$^-$  \\
          560.5             & 506.5       &19(4)                &0.90(10)$^e$       &               & E2            & 13/2$^-$ $\rightarrow$ ~9/2$^-$  \\
          585.1             & 1520.4     & 7.3(9)              &0.55(3)$^e$         &               & M1/E2      & 21/2$^-$ $\rightarrow$ 19/2$^-$  \\  
          594.3             & 793.1       & 62(4)               &1.00(8)$^e$         &               & E2            & 15/2$^-$ $\rightarrow$ 11/2$^-$     \\
          667.7             & 3043.3     & 0.8(3)              &0.82(9)$^e$         &               & M1/E2      & 25/2$^-$ $\rightarrow$ 23/2$^-$  \\
          709.3             & 1269.8     & 24(3)               &1.08(8)$^e$         &               & E2            & 17/2$^-$ $\rightarrow$  13/2$^-$ \\
          727.3             & 1520.4     & 37.0(35)          &1.08(9)$^e$         &               & E2            & 19/2$^-$ $\rightarrow$ 15/2$^-$    \\
          748.2             & 4067.7     &1.12(6)             &0.65(3)$^e$         &               & M1/E2      & 29/2$^-$ $\rightarrow$ 27/2$^-$  \\
          835.7             & 2105.5     & 19(5)               &0.90(13)$^e$       &               & E2            & 21/2$^-$ $\rightarrow$ 17/2$^-$     \\
          855.2             & 2375.6     & 21.2(20)          &1.15(17)$^e$       &               & E2            & 23/2$^-$ $\rightarrow$ 19/2$^-$  \\   
          937.8             & 3043.3     & 2.3(5)              &1.06(11)$^e$       &               & E2            & 25/2$^-$ $\rightarrow$ 21/2$^-$     \\     
          943.9             & 3319.5     & 3.5(5)              &1.12(20)$^e$       &               & E2            & 27/2$^{-}$ $\rightarrow$ 23/2$^{-}$ \\    
          (1024.4)         & 4067.7     &                         &                            &               &                 &(29/2$^-$)$\rightarrow$27/2$^-$  \\
          1032.3           & 4351.8     & 1.2(1)              & 0.83(23)$^e$      &               & E2            & 31/2$^-$ $\rightarrow$ 27/2$^-$    \\       
 \\                      
 $\bf Band ~D2 $ & & &         &            &                                      &          \\
         177.2               & 1954.5       & 2.0(1)               &                          &0.66(7)       & M1/E2        & 17/2$^+$ $\rightarrow$ 15/2$^+$  \\
         192.4               & 2350.8       & 17(2)                &0.64(3)$^e$       &                   & M1/E2        & 21/2$^+$ $\rightarrow$ 19/2$^+$  \\
          203.9              &  2158.4      & 9.5(8)               &0.85(9)$^e$       &                   & M1/E2        & 19/2$^+$ $\rightarrow$ 17/2$^+$    \\
          207.5              & 2588.3       & 13.5(25)           &0.65(7)$^e$       &                   & M1/E2        & 23/2$^+$ $\rightarrow$ 21/2$^+$    \\
          243.2              &  2801.5      & 9.2(9)               &0.64(9)$^e$       &                   & M1/E2        & 25/2$^+$ $\rightarrow$ 23/2$^+$  \\
          301.1              & 3102.6       & 7.5(6)               &0.34(7)$^e$       &                        & M1/E2        & 27/2$^+$ $\rightarrow$ 25/2$^+$    \\
          369.1              &  3471.7      & 4.9(4)               &0.82(36)$^e$     &                        & M1/E2        & 29/2$^+$ $\rightarrow$ 27/2$^+$  \\
          381.1              & 2158.4       & 0.55(7)             &                          &1.32(23)           & E2              & 19/2$^+$ $\rightarrow$ 15/2$^+$  \\
          390.0              & 3861.7       & 3.6(4)               &                          &0.65(5)             & M1/E2        & 31/2$^+$ $\rightarrow$ 29/2$^+$    \\
          396.3              & 2350.8       & 1.29(21)           &                          &1.37(17)           & E2              & 21/2$^+$ $\rightarrow$ 17/2$^+$    \\
          397.7              & 6550.0       & 0.4(1)            &                          &0.79(11)            & M1/E2        & 43/2$^+$ $\rightarrow$ 41/2$^+$    \\
          399.9              & 2588.3       & 0.82(20)            &                          &1.36(16)            & E2              & 23/2$^+$ $\rightarrow$ 19/2$^+$  \\
          416.7              &  6152.3      & 0.6(1)             &                          &0.86(20)            & M1/E2        & 41/2$^+$ $\rightarrow$ 39/2$^+$  \\
          421.9              & 5735.6       & 0.31(11)             &                          &1.09(16)            & M1/E2        & 39/2$^+$ $\rightarrow$ 37/2$^+$    \\
          425.4              & 4773.2       & 1.20(21)             &                          &1.07(14)            & M1/E2        & 35/2$^+$ $\rightarrow$ 33/2$^+$    \\
          436.3              & 6986.3       & 0.20(3)             &                          &                         &                    &(45/2$^+$)$\rightarrow$43/2$^+$  \\
          450.7              &  2801.5      & 0.86(12)           &                          & 1.38(11)           & E2              & 25/2$^+$ $\rightarrow$ 21/2$^+$     \\
          486.1              &  4347.8      & 2.10(25)            &                          &0.75(2)              & M1/E2        & 33/2$^+$ $\rightarrow$ 31/2$^+$  \\
          540.5              &  5313.7      & 0.85(15)           &                          &1.22(10)            & M1/E2        & 37/2$^+$ $\rightarrow$ 35/2$^+$  \\
          544.3              & 3102.6       &1.10(25)            &1.08(12)$^e$     &                         & E2              & 27/2$^+$ $\rightarrow$ 23/2$^+$  \\
          670.2             &  3471.7       &1.43(22)            &0.95(16)$^e$     &                         & E2              & 29/2$^+$ $\rightarrow$ 25/2$^+$     \\
          759.1             & 3861.7        &1.7(3)                &                          &1.42(12)            & E2              & 31/2$^+$ $\rightarrow$ 27/2$^+$  \\
          814.4             & 6550.0        & 0.20(5)           &                          &1.36(10)            & E2              & 43/2$^+$ $\rightarrow$ 39/2$^+$  \\
          830.4             & 2350.8        & 5.0(3)               &0.48(5)$^e$       &                          & E1              & 21/2$^+$ $\rightarrow$ 19/2$^-$  \\
          834.0             &  6986.3       &0.10(4)            &                          &                         &                    &(45/2$^+$)$\rightarrow$41/2$^+$     \\
          838.6             &  6152.3       & 0.27(7)              &                          & 1.34(21)           & E2              & 41/2$^+$ $\rightarrow$ 37/2$^+$     \\
          876.1             &  4347.8       & 1.20(17)          &                          &1.50(17)            & E2              & 33/2$^+$ $\rightarrow$ 29/2$^+$     \\
          888.6             &  2158.4       & 8.1(6)               &0.58(15)$^e$     &                          & E1              & 19/2$^+$ $\rightarrow$ 17/2$^-$  \\
          911.5             & 4773.2        & 1.7(3)               &                          &1.41(15)            & E2              & 35/2$^+$ $\rightarrow$ 31/2$^+$  \\
          962.4             & 5735.6        & 0.7(2)               &                          &1.43(16)            & E2              & 39/2$^+$ $\rightarrow$ 35/2$^+$  \\
          965.9             &  5313.7       &1.2(2)                &                          &1.37(25)            & E2              & 37/2$^+$ $\rightarrow$ 33/2$^+$     \\
         1161.4           & 1954.5        & 11(1)                 &0.60(16)$^e$    &                          & E1              & 17/2$^+$ $\rightarrow$ 15/2$^-$     \\
         1216.8           & 1777.3        & 3.2(2)               &                          &0.83(10)            & E1              & 15/2$^+$ $\rightarrow$ 13/2$^-$  \\        
  \\
 $\bf Band~D3 $ & & &         &            &                                                \\
          191.5            & 2535.0        & 1.4(2)                &                           &  1.10(16)           & M1/E2          & 21/2$^+$ $\rightarrow$ 19/2$^+$  \\
          235.7            & 2770.7        & 3.4(2)                 &0.62(11)$^f$      &                           & M1/E2          & 23/2$^+$ $\rightarrow$ 21/2$^+$     \\
          285.0            & 3055.7       & 4.0(3)                 &0.35(8)$^f$        &                           & M1/E2          & 25/2$^+$ $\rightarrow$ 23/2$^+$    \\
          328.5            &  3718.6       & 1.1(1)                 &0.87(7)$^f$        &                           & M1/E2          & 29/2$^+$ $\rightarrow$ 27/2$^+$  \\
          334.4            & 3390.1        & 2.1(1)                 &0.61(14)$^f$      &                           &  M1/E2         & 27/2$^{-}$ $\rightarrow$ 25/2$^{+}$ \\
          376.6            & 2535.0       & 0.17(2)               &                          &                             &                     & 21/2$^+$ $\rightarrow$ 19/2$^+$  \\
         389.0            & 2343.5       & 0.43(6)               &                          &                             &                     & 19/2$^+$ $\rightarrow$ 17/2$^+$  \\
         402.9            &  4540.8      & 0.30(4)               &                          &                             &                     & 33/2$^+$ $\rightarrow$ 31/2$^+$  \\
         419.3            & 4137.9        & 0.39(5)               &                          &1.10(25)              & M1/E2          & 31/2$^+$ $\rightarrow$ 29/2$^+$  \\
         419.9            & 2770.7       & 0.44(4)               &                          &1.04(16)                & M1/E2          & 23/2$^+$ $\rightarrow$ 21/2$^+$    \\
         520.3            & 4658.2        & 0.13(2)               &                          &                            &                     &(33/2$^+$)$\rightarrow$31/2$^+$  \\
         566.2            & 2343.5       & 0.25(2)               &0.85(25)$^e$     &                             & E2                & 19/2$^+$ $\rightarrow$ 15/2$^+$    \\
         580.5            & 2535.0       & 0.9(1)                 &                          &1.34(21)                & E2               & 21/2$^+$ $\rightarrow$ 17/2$^+$    \\
         588.6            & 3390.1       & 0.26(2)               &                          &                              &                    & 27/2$^+$ $\rightarrow$ 25/2$^+$  \\
         612.3            & 2770.7       & 0.52(7)              &                          &1.44(23)               & E2                & 23/2$^+$ $\rightarrow$ 19/2$^+$  \\
         616.0            & 3718.6       &1.19(11)              &                          &1.02(14)               & M1/E2          & 29/2$^+$ $\rightarrow$ 27/2$^+$  \\
         662.9            &  3718.6       & 0.5(2)                 &                          &                            &                     & 29/2$^+$ $\rightarrow$ 25/2$^+$     \\
         666.2            & 4137.9       & 0.58(5)               &                          &0.65(8)                 & M1/E2         & 31/2$^+$ $\rightarrow$ 29/2$^+$  \\
         679.1            &  4540.8      & 0.43(2)               &                          &0.58(4)                 & M1/E2         & 33/2$^+$ $\rightarrow$ 31/2$^+$  \\
         704.9            & 3055.7       & 0.57(2)               &1.11(16)$^e$     &                            & E2               & 25/2$^+$ $\rightarrow$ 21/2$^+$  \\
         831.8            & 3390.1       & 0.36(2)               &1.2(3)$^e$         &                            & E2               & 27/2$^+$ $\rightarrow$ 23/2$^+$     \\ 
         917.1            &  3718.6      & 0.26(3)               &                          &                            &               & 29/2$^+$ $\rightarrow$ 25/2$^+$    \\
         1014.6           & 2535.0       & 1.7(1)                &0.51(6)$^e$       &                            & E1               & 21/2$^+$ $\rightarrow$ 19/2$^-$     \\
         1035.3           & 4137.9       & 0.19(2)              &                          &1.30(18)               & E2               & 31/2$^+$ $\rightarrow$ 27/2$^+$    \\
         1069.1           &  4540.8      & 0.29(2)              &                          &1.43(15)               & E2               & 33/2$^+$ $\rightarrow$ 29/2$^+$     \\
         1073.7           & 2343.5       & 1.2(1)                &0.64(25)$^e$     &                            & E1               & 19/2$^+$ $\rightarrow$ 17/2$^-$  \\
 \\  
$\bf Band~D4$ & &            &                 &                  &         \\  
          197.3            &  2704.3       & 0.40(5)               &                        &                             &                     & 21/2$^+$ $\rightarrow$ 19/2$^+$    \\
          235.7            & 2940.0        & 1.7(2)                 &0.31(3)$^e$     &                             & M1/E2          & 23/2$^+$ $\rightarrow$ 21/2$^+$  \\
          267.6            &  3207.6       & 1.9(2)                 &0.22(6)$^e$     &                             & M1/E2          & 25/2$^+$ $\rightarrow$ 23/2$^+$     \\
          343.5            &  3551.1       & 1.2(1)                 &0.44(7)$^e$     &                             &M1                & 27/2$^+$ $\rightarrow$ 25/2$^+$  \\
          423.4            &  3974.5       & 0.25(3)               &                        &1.08(16)                & M1/E2          & 29/2$^+$ $\rightarrow$ 27/2$^+$  \\    
          589.2            & 2940.0        & 0.21(3)               &                        &1.45(30)                & E2               & 23/2$^+$ $\rightarrow$ 19/2$^+$  \\
          596.5            & 2940.0        & 0.13(5)               &                        &                             &                     & 23/2$^+$ $\rightarrow$ 19/2$^+$    \\
          649.3            & 3207.6        & 0.35(3)               &                         & 1.41(17)              & E2               & 25/2$^+$ $\rightarrow$ 21/2$^+$  \\
          672.6            &  3207.6       & 0.11(2)               &                         &                             &                    & 25/2$^+$ $\rightarrow$ 21/2$^+$  \\
          749.8            &  2704.3       & 0.15(2)               &                         &                            &                     & 21/2$^+$ $\rightarrow$ 17/2$^+$  \\
          780.4            &  3551.1       & 0.13(4)               &0.90(20)$^e$   &                            & E2               & 27/2$^+$ $\rightarrow$ 23/2$^+$  \\
          918.8            &  3974.5       & 0.21(5)               &                         &                            &                     & 29/2$^+$ $\rightarrow$ 25/2$^+$  \\
         1183.9           &  2704.3       & 1.3(1)                 &0.61(16)$^e$    &                            & E1                & 21/2$^+$ $\rightarrow$ 19/2$^-$     \\
         1237.2           & 2507.0        &0.45(4)                &                         & 0.72(15)              & E1               & 19/2$^+$ $\rightarrow$ 17/2$^-$  \\
\\
  $\bf Band~D5$ & &            &                 &                      &      &   \\
         121.0              & 2940.6       & 6.0(5)               &0.41(5)$^e$       &                            & M1             & 25/2$^{-}$ $\rightarrow$ 23/2$^{-}$ \\
         129.8              & 2819.6       & 0.12(2)             &                          &                            &                    & 23/2$^-$ $\rightarrow$ 21/2$^{-}$ \\
         170.4              & 3111.0        & 24.5(20)          &0.44(3)$^e$       &                             &M1              & 27/2$^-$ $\rightarrow$ 25/2$^-$     \\ 
          247.9             & 3358.9       & 18.8(17)           &0.55(9)$^e$       &                            & M1/E2        & 29/2$^{-}$ $\rightarrow$ 27/2$^{-}$ \\
          250.8             & 2940.6       & 0.16(3)             &                          &                            &                    & 25/2$^-$ $\rightarrow$  21/2$^-$ \\
          291.1             & 3650.0       & 12.7(14)           &0.53(7)$^e$       &                            & M1/E2         & 31/2$^-$ $\rightarrow$ 29/2$^-$  \\
          291.4             & 3111.0       & 0.61(20)            &                          &                            &                    & 27/2$^-$ $\rightarrow$ 23/2$^-$    \\ 
          358.2             & 4008.2       & 9.9(6)               &0.58(15)$^e$     &                            & M1/E2         & 33/2$^{-}$ $\rightarrow$ 31/2$^{-}$ \\
          406.3             & 4414.5       & 6.6(5)               &0.63(5)$^e$       &                            & M1/E2         & 35/2$^-$ $\rightarrow$ 33/2$^-$  \\
          418.3             & 3358.9       & 1.0(2)               &0.94(9)$^e$       &                            & E2               & 29/2$^-$ $\rightarrow$  25/2$^-$ \\
          438.2             & 4852.7       & 5.2(5)               &0.43(3)$^e$       &                            & M1              & 37/2$^-$ $\rightarrow$ 35/2$^-$    \\  
          444.0             & 2819.6       & 1.7(1)              & 1.22(19)$^e$    &                            & M1/E2           & 23/2$^-$ $\rightarrow$ 23/2$^-$   \\
          462.4             & 5315.2       & 3.9(3)               &0.70(5)$^e$       &                            & M1/E2         & 39/2$^-$ $\rightarrow$  37/2$^-$ \\
          472.4             & 5787.6       & 2.5(2)               &                          &0.79(6)                 & M1              & 41/2$^-$ $\rightarrow$ 39/2$^-$    \\    
          493.9            & 6281.5      & 1.53(15)             &                           &1.11(12)               & M1/E2         & 43/2$^-$ $\rightarrow$  41/2$^-$ \\
          539.0            & 3650.0      & 1.7(2)                 &1.04(12)$^e$       &                           & E2               & 31/2$^-$ $\rightarrow$ 27/2$^-$    \\  
          560.1            & 6841.6      & 0.85(7)               &                           &                            &                    & 45/2$^-$ $\rightarrow$ 43/2$^-$    \\
          565.0             & 2940.6      & 8.4(6)               &0.37(5)$^e$        &                            & M1/E2          & 25/2$^-$ $\rightarrow$ 23/2$^-$     \\
          565.0            & 7406.6      & 0.5(1)                 &                           &                            &                    &(47/2$^-$)$\rightarrow$45/2$^-$ \\ 
          584.3             & 2689.6     & 0.50(4)             &0.48(5)$^e$        &                            &M1                & 21/2$^-$ $\rightarrow$  21/2$^-$ \\
          640.6             & 2819.6      & 0.11(2)             &                           &                            &                      & 23/2$^-$ $\rightarrow$ 19/2$^-$  \\
          649.3             & 4008.2      & 1.82(20)            &0.89(11)$^e$      &                            & E2                & 33/2$^-$ $\rightarrow$  29/2$^-$ \\
          714.1             & 2819.6      & 2.9(2)               &0.68(7)$^e$        &                            & M1/E2          & 23/2$^-$ $\rightarrow$ 21/2$^-$  \\
          735.4             & 3111.0       & 2.9(2)              &0.97(12)$^e$       &                           &E2                  & 27/2$^-$ $\rightarrow$ 23/2$^-$  \\
          764.5             & 4414.5      & 2.4(3)                &1.23(26)$^e$       &                           & E2                 & 35/2$^-$ $\rightarrow$ 31/2$^-$    \\  
          835.1             & 2940.6      & 17.9(18)          &1.02(13)$^e$       &                           & E2                  & 25/2$^-$ $\rightarrow$ 21/2$^-$  \\ 
          844.5             & 4852.7      & 2.6(3)               &                            &1.34(15)              & E2                 & 37/2$^-$ $\rightarrow$ 33/2$^-$  \\
          900.7             & 5315.2      & 0.95(20)           &                            &1.51(19)               & E2                 & 39/2$^{-}$ $\rightarrow$ 35/2$^{-}$ \\
          934.9             & 5787.6      & 0.91(15)           &                            &1.46(23)              & E2                 & 41/2$^-$ $\rightarrow$ 37/2$^-$  \\
          966.3             & 6281.5      & 0.68(10)           &                          &1.39(17)              & E2                 & 43/2$^{-}$ $\rightarrow$ 39/2$^{-}$ \\
          1054.0           & 6841.6      & 0.52(10)           &                            &1.64(33)              & E2                 & 45/2$^-$ $\rightarrow$ 41/2$^-$  \\
          1125.1           & 7406.6      & 0.45(5)             &                            &                            &                      &(47/2$^{-}$)$\rightarrow$43/2$^{-}$ \\
          1299.2           & 2819.6      & 0.76(4)            &1.14(16)$^e$       &                           & E2                  & 23/2$^{-}$ $\rightarrow$ 19/2$^{-}$ \\
\\     
          $\bf Band~D6$ &&              &                &                 &          &  \\
          173.0            & 3782.0     & 2.3(2)              &0.7(2)$^e$             &                           & M1/E2           & 29/2$^{-}$ $\rightarrow$ 27/2$^{-}$ \\
          225.8            & 4007.8     & 3.8(4)              &0.85(9)$^e$           &                           & M1/E2           & 31/2$^{-}$ $\rightarrow$ 29/2$^{-}$ \\
          282.4            & 4290.2     & 4.5(3)              &                              &0.61(9)                & M1/E2           & 33/2$^{-}$ $\rightarrow$ 31/2$^{-}$ \\
          309.4            & 4599.6     & 3.1(2)              &                              &0.66(7)                & M1/E2           & 35/2$^{-}$ $\rightarrow$ 33/2$^{-}$ \\
          371.6            & 4971.2     & 2.5(5)              &                              &                           &                       & 37/2$^-$ $\rightarrow$ 35/2$^-$ \\  %
          440.4            & 5411.6     & 0.95(17)          &                              &1.09(22)              & M1/E2           & 39/2$^{-}$ $\rightarrow$ 37/2$^{-}$ \\ %
          463.0            & 3782.0      & 1.12(13)          &                              &0.67(8)                & M1/E2           & 29/2$^-$ $\rightarrow$ 27/2$^-$  \\
          512.4            & 5924.0     & 0.20(6)            &                              &                           &                      & (41/2$^{-}$) $\rightarrow$ 39/2$^{-}$ \\
          591.4            & 4599.6      & 0.8(2)              &                              &                            &            & 35/2$^-$ $\rightarrow$ 33/2$^-$    \\
          591.8           & 4599.6      & 0.6(2)              &                             &                            &                      & 35/2$^-$ $\rightarrow$ 31/2$^-$  \\
          608.8            & 5924.0      & 0.25(4)            &                              &                            &          & (41/2$^-$ )$\rightarrow$ 39/2$^-$    \\
          640.2            & 4290.2      & 0.57(3)            &                              &1.38(23)              & M1/E2            & 33/2$^-$ $\rightarrow$ 31/2$^-$    \\
          648.9            & 4007.8      & 1.8(2)              &                              &0.22(3)                 & M1/E2            & 31/2$^-$ $\rightarrow$ 29/2$^-$    \\
          668.4            & 3609.0      &2.5(2)               &0.66(12)$^e$         &                            & M1/E2            & 27/2$^-$ $\rightarrow$ 25/2$^-$  \\
          671.0            & 3782.0      & 2.1(2)              &0.73(16)$^e$         &                            & M1/E2            & 29/2$^-$ $\rightarrow$ 27/2$^-$    \\  
          681.0           & 4971.2      & 0.55(10)          &                             &                            &                       &  37/2$^{-}$$\rightarrow$33/2$^{-}$ \\   
          812.0           & 5411.6      & 0.18(5)            &                             &                            &                       &39/2$^{-}$$\rightarrow$35/2$^{-}$ \\  
          841.4            & 3782.0      & 1.58(17)          &0.55(9)$^e$           &                            & M1/E2            & 29/2$^-$ $\rightarrow$  27/2$^-$ \\
          896.8            & 4007.8      & 0.79(6)            &1.02(16)$^e$       &                              & E2            & 31/2$^-$ $\rightarrow$  27/2$^-$ \\
          931.3            & 4290.2      & 0.50(3)            &                            &1.53(17)                 & E2            & 33/2$^-$ $\rightarrow$  29/2$^-$ \\
          934.9             & 5787.6     & 0.70(20)          &                            &1.46(23)              & E2                 & 41/2$^-$ $\rightarrow$ 37/2$^-$  \\
          949.6            & 4599.6      & 0.69(8)            &                            &1.51(27)                 & E2            & 35/2$^-$ $\rightarrow$  31/2$^-$ \\        
          963.0            & 4971.2      & 0.53(5)            &                            &1.47(14)                 & E2            & 37/2$^-$ $\rightarrow$ 33/2$^-$    \\
          966.3             & 6281.5     & 0.68(10)          &                            &1.39(17)              & E2                 & 43/2$^{-}$ $\rightarrow$ 39/2$^{-}$ \\
          998.1            & 5412.6      & 0.85(6)            &                            &1.42(16)                 & E2             & 39/2$^-$ $\rightarrow$  35/2$^-$ \\
          1071.3          & 5924.0      & 0.22(3)            &                            &                              &             & (41/2$^-$) $\rightarrow$  37/2$^-$ \\
\\
\hline 
\end{longtable*}
$^a$The error on the transition energies is 0.2 keV for transitions below 1000 keV of the $^{135}$Nd reaction channel, 0.5 keV for transitions above 1000 keV and 1 keV for transitions above 1200 keV. \\

$^b$Relative intensities corrected for efficiency, normalized to the intensity of the 198.8 keV transition. The transition intensities were obtained from a combination of total projection and gated spectra.  \\

$^c$$R_{DCO}$ has been deduced from asymmetric $\gamma$-$\gamma$ coincidence matrix sorted with detectors at $157.6^{\circ}$ on one axis, and detectors at $\approx$ $90^{\circ}$ on the other axis. The tentative spin - parity of the states are given in parenthesis. \\

$^d$$R_{ac}$ has been deduced from two asymmetric $\gamma$-$\gamma$ coincidence matrices sorted with detectors at $133.6^{\circ}$ and $157.6^{\circ}$ on one axis, and detectors at $\approx $ $90^{\circ}$ on the other axis, respectively. The tentative spin - parity of the states are given in parenthesis. \\

$^e$DCO ratio from spectrum gated on stretched quadrupole transition. \\

$^f$DCO ratio from spectrum gated on stretched dipole transition.\\ 

\end{document}